%% file: ms.tex
\documentclass[manuscript]{aastex}

\slugcomment{to be submitted for the regular issue of the Astrophysical Journal}

\shorttitle{Long term variation of the solar diurnal anisotropy}
\shortauthors{Munakata et al.}

\begin{document}

\title{Long term variation of the solar diurnal anisotropy of galactic cosmic rays observed with the Nagoya multi-directional muon detector}

\author{K. Munakata, M. Kozai, C. Kato}
\affil{Physics Department, Shinshu University, Matsumoto, Nagano 390-8621, Japan}
\email{kmuna00@shinshu-u.ac.jp}
\and
\author{J. K\'ota}
\affil{Lunar and Planetary Laboratory, University of Arizona, Tucson, AZ 87721, USA}

\begin{abstract}
We analyze the three dimensional anisotropy of the galactic cosmic ray (GCR) intensities observed independently with a muon detector (MD) at Nagoya in Japan and neutron monitors over four solar activity cycles. We clearly see the phase of the free-space diurnal anisotropy shifting toward earlier hours around solar activity minima in $A>0$ epochs, due to the reduced anisotropy component parallel to the mean magnetic field. This component is consistent with a rigidity independent spectrum, while the perpendicular anisotropy component increases with GCR rigidity. We suggest that this harder spectrum of the perpendicular component is due to contribution from the drift streaming. We find that the bidirectional latitudinal density gradient is positive in $A>0$ epoch, while it is negative in $A<0$ epoch, in accord with the drift model prediction. The radial density gradient of GCRs, on the other hand, varies with $\sim$11-year cycle with maxima (minima) in solar maximum (minimum) periods, but we find no significant difference between the radial gradients in $A>0$ and $A<0$ epochs. The corresponding parallel mean free path is larger in $A<0$ than in $A>0$. We also find, however, that parallel mean free path (radial gradient) appears to persistently increase (decreasing) in the last three cycles of weakening solar activity. We suggest that simple differences between these parameters in $A>0$ and $A<0$ epochs are seriously biased by these long-term trends.
\end{abstract}


\section{Introduction}
The solar wind is a supersonic plasma blowing radially outward from the sun toward a vast space filled by cold and thin interstellar plasma. The global structure of the region called the ``heliosphere'', which is a region dominated by the solar wind plasma and the solar magnetic field, is of great interest for both solar- and astrophysicists. The Interplanetary Magnetic Field (IMF) is the term representing the solar magnetic field carried outward by the solar wind into the heliosphere as magnetic field lines from the Sun are dragged along by the highly conductive solar wind plasma \citep{Par58}. Because of the dominant dipole component of the solar magnetic field, the IMF is divided into two magnetic sectors in the northern and southern hemisphere separated by the Heliospheric Current Sheet (HCS) which develops into a ``wavy'' three dimensional structure. The inclination of the magnetic dipole from the rotation axis increases with increasing the solar activity and reverses during the solar activity maximum epoch when the inclination becomes maximum. The Sun has a strong and complex magnetic field, and the physical properties of the heliosphere is directly connected to the properties of the magnetic field varying with a period of about 11 years.\par
Temporal variations in the inner heliosphere can be deduced from the ground-based observations of the high-energy Galactic Cosmic Rays (GCRs). GCRs are high-energy nuclei (mostly protons) accelerated in our galaxy and continuously arriving at the earth after traveling through the heliosphere. After entering the heliosphere, GCRs interact with the IMF being carried outward by the solar wind. The interaction with the large-scale ordered field causes the gradient- and curvature-drift motions of GCRs in the heliosphere, while the interaction with the irregular (or disordered) field component results in the pitch angle scattering of GCRs. The scattering by the magnetic irregularities embedded in the expanding solar wind causes the deceleration (called the adiabatic cooling) and also causes an outward convection which leads to lower GCR intensities closer to the Sun. The resulting positive radial gradient of GCRs produces an inward diffusion, flowing preferentially along the ordered IMF lines. A steady state distribution is realized when the inward diffusion is balanced with the outward convection. The GCR intensity measured at the Earth changes with various time scales. The solar cycle variation of the solar wind parameters, such as the solar wind velocity, the magnitude and orientation of the IMF, the tilt angle of the HCS and the mean free path of the pitch angle scattering of GCRs in the turbulent magnetic field, alters the spatial distribution of GCR density in the heliosphere. The drift model of GCR transport predicts a bi-directional latitudinal gradient pointing in opposite directions on the opposite sides of the HCS if the HCS is flat \citep{Jok79}. The predicted spatial distribution of the GCR density has a minimum along the HCS in the ``positive'' polarity period of the solar polar magnetic field (also referred as $A>0$ epoch), when the IMF directs away from (toward) the Sun in the northern (southern) hemisphere, while the distribution has the local maximum on the HCS in the ``negative'' period ($A<0$ epoch) with the opposite field orientation in each hemisphere. The field orientation reverses every 11 years around the maximum period of the solar activity. A tilted current sheet introduces modifications around the wavy HCS. For example the intensity minimum (for $A>0$) will not be right at the HCS, but the general tendencies in the sense of the latitudinal gradient remain the same as outlined above \citep{Jok82}.\par
The variation of the spatial distribution of GCR density causes the variation of the directional anisotropy of the GCR intensity measured at the Earth. One such variation is the 22-year variation of the solar diurnal anisotropy in which the phase (or the local solar time of maximum intensity) of the anisotropy shifts towards earlier hours around every $A>0$ solar minima \citep[][and references therein]{Tha53, For67, Ahl88, Bie91}. By analyzing the anisotropy observed with neutron monitors (NMs) in 1968-1988, \citet{Che93} (hereafter referred as Paper I) revealed that the observed phase-shift of the diurnal anisotropy is due to the decrease of the diffusion streaming parallel to the IMF in $A>0$ solar minima. The parallel diffusion streaming is proportional to the radial gradient ($G_r$) of GCR density multiplied by the parallel mean free path ($\lambda_{\|}$) of the pitch angle scattering. The simple drift model predicts smaller $G_r$ in $A>0$ epoch than in $A<0$ epoch, if the diffusion coefficients are same in both epochs \citep{Kot83}. Finding a significant 11-year solar cycle variation but no clear 22-year variation in the observed $G_r$, however, Paper I suggested that the smaller parallel streaming in the $A>0$ solar minimum period was caused by the smaller $\lambda_{\|}$, possibly due to the magnetic helicity effect in the turbulent magnetic field \citep[Paper I;][]{Bie86, Bie87}.\par
The GCR anisotropy (or the streaming) vector in three dimensions (3D) consists of three components, two lying in the ecliptic plane and the other pointing normal to the ecliptic plane. The two ecliptic components, parallel and perpendicular to the IMF, are derived from the amplitude and phase of the solar diurnal anisotropy corrected for the contribution from the radial solar wind convection. Paper I analyzed the diurnal anisotropy in free space, corrected for the geomagnetic deflection of GCR orbits, by assuming a power law type ($\propto p^\gamma$) dependence of the anisotropy amplitude on the GCR rigidity ($p$) with the spectral index ($\gamma$) and the upper limiting rigidity ($P_u$) fixed at 0 and 100 GV, respectively. The zero spectral index of the diurnal anisotropy has been assumed in many analyses based on the original convection-diffusion picture of the GCR transport in which the stationary GCR distribution in the heliosphere results from the inward diffusion balancing with the outward convection by the solar wind which is independent of the rigidity \citep{Par65, Gle67, Gle69}. The upper limiting rigidity ($P_u$) set at 100 GV was also a reasonable assumption for the analysis of NM data alone, because $P_u$ representing the break-down rigidity of the diffusion picture is expected to be much higher than the median primary rigidity to which the NMs used in Paper I respond. \citet{Mun97} assumed $\gamma=0$ but treated $P_u$ as a free parameter in their analyses of the diurnal anisotropy observed with multi-directional muon detectors (MDs) which have median responses to GCRs with higher rigidity than NMs. They found $P_u$ changing between 100 and 300 GV in a clear correlation with the solar activity \citep{Mun02}. \citet{Hal97} treated both $\gamma$ and $P_u$ as free parameters in their analyses of the NM and MD data and reported the temporal variation of each parameter in solar activity and solar magnetic cycles.\par
All these works take account of the rigidity dependence of the amplitude varying as a function of time, but they still assume that the phase is independent of rigidity. In other words, they assumed a common rigidity spectrum for two ecliptic components, parallel and perpendicular to the IMF. \citet{Bie91} (hereafter referred as Paper II), on the other hand, also reported that the magnitude of the observed phase variation in $A>0$ solar minimum increases with GCR rigidity \citep{Agr83}. This rigidity dependent feature of the observed phase variation cannot be reproduced properly, as long as the rigidity spectrum common for two ecliptic components is assumed. This observed feature has been confirmed by other papers \citep[e.g.][]{Oh10}, but its physical origin is still left unknown.\par
The third component of the anisotropy, that is, the north-south (NS) anisotropy normal to the ecliptic plane has been derived also from NM and MD data in a couple of different ways. \citet{Bie86} and Paper I derived this anisotropy from the difference between count rates in a pair of NMs which are located near the north and south geomagnetic poles and observing intensities of GCRs arriving from the north and south pole orientations, respectively. They found a $\sim$10-year cycle variation in this component anisotropy which implied the radial gradient ($G_r$) of GCR density changing in correlation with the solar activity, while they found no significant difference of $G_r$ in $A>0$ and $A<0$ epochs in a contradiction with the simple drift model prediction. Due to 23.4$\degr$ inclination of Earth's rotation axis from the ecliptic normal, the NS anisotropy normal to the ecliptic plane can be also observed as a diurnal variation of count rate in the sidereal time with the maximum phase at $\sim$18:00 local sidereal time \citep{Swi69}. \citet{Yas80} analyzed this sidereal diurnal variation observed by NMs and MDs during 5 years between 1969 and 1973 and found that observations were reproduced best by the average rigidity spectrum with $\gamma=0.3$ and $P_u=200$ GV. This was the first experimental indication that the rigidity spectrum of the anisotropy has a positive spectral index. \citet{Hal94} also applied the same method to NM and MD data observed between 1957 and 1985 and found the average spectrum with $\gamma=0.5$ and $P_u=400$ GV, again with a positive $\gamma$. This suggested that each of two ecliptic components may also have a spectrum with non-zero $\gamma$.\par
A possible drawback of deriving the NS anisotropy from the sidereal diurnal variation is that the expected amplitude of the sidereal diurnal variation ($\sim$0.03 \%) is approximately an order of magnitude smaller than the solar diurnal variation ($\sim$0.3 \%). The small signal in the sidereal time can be easily influenced by the solar diurnal anisotropy changing in a year. Another difficulty is that one can obtain only the yearly mean anisotropy. This is because of the fact that the influence from the solar diurnal variation, even if it is stationary through a year, can be eliminated in the sidereal time only when the diurnal variation is averaged over integral year(s). This makes it difficult to deduce reliable error of the yearly mean anisotropy. \citet{Mor79} proposed another way to derive the NS anisotropy from the ``GG-component'' of a multi-directional MD at Nagoya in Japan. The GG-component is a difference combination between intensities recorded in the north- and south-viewing directional channels designed to measure the NS anisotropy free from the atmospheric temperature effect \citep{Nag72}. \citet{Lau03} showed that GG-component can be used for deriving reliable sector polarity of the IMF. By using a global network of four multi-directional MDs which are capable of observing the NS anisotropy on hourly basis, \citet{Oka08} confirmed that the north-south anisotropy deduced from the GG-component is consistent with the anisotropy observed with the global network.\par
In the present paper, we extend the analysis by Paper I to the most recent period and derive the long-term variation of the modulation parameters from the 3D anisotropy observed during 44 years by the Nagoya multi-directional MD which has a median rigidity of 60 GV for primary GCRs. We also analyze the anisotropy observed during the same period by NMs which have the median response to 17 GV primary GCRs. We derive the NS anisotropy from the GG-component of the Nagoya MD. We particularly examine the rigidity dependences of each component of the anisotropy and each modulation parameter by comparing them derived from MD and NM data at 60 GV and 17 GV, respectively. We do not intend to determine each rigidity spectrum quantitatively by, for instance, calculating both $\gamma$ and $P_u$ as free parameters in best-fit calculation as a function of time. In such best-fit calculations, we often see a significant anti-correlation between the best-fit $\gamma$ and $P_u$ \citep{Hal94, Hal97}. A large $P_u$ with a small (or negative) $\gamma$ often returns similar $\chi^2$-value as a small $P_u$ with large (or positive) $\gamma$ does, increasing the systematic error of each best-fit value. We instead examine the rigidity spectrum qualitatively based on the ratio between parameters derived from NM and MD data with a common assumption of the spectrum with fixed values of $\gamma =0$ and $P_u=100$ GV, respectively, as done in Paper I. If the ratio is close to one, the spectrum is consistent with the assumption. If the ratio is significantly larger (smaller) than one, on the other hand, we can conclude that the spectrum is harder (softer) than the assumed one. In this way, we can make a qualitative but reliable examination of the rigidity dependence of each parameter. We will present quantitative analyses of the rigidity dependence elsewhere. It will be shown in the present paper that three components of the anisotropy have different rigidity dependence. This naturally explains the rigidity dependent feature of the observed phase variation mentioned above. We will also suggest that the different rigidity dependence for three anisotropy components are possibly due to the relative contribution from the drift (diamagnetic drift) which is different in each component.\par
The outline of this paper is as follows. In section 2, we describe the data analysis and results in detail. The conclusions and discussions are given in section 3. For readers' references, we also present our results in a numerical data table in Appendix A. In Appendix B, we show how the obtained results depend on the assumed value of $P_u$.

\section{Data analyses and results}
\begin{figure}
\plotone{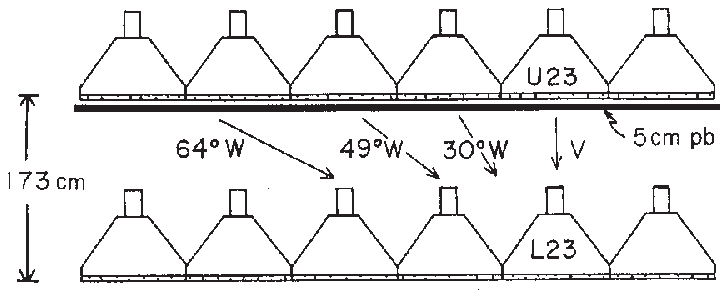}
\caption{Nagoya multi-directional muon detector. This figure is supplied from a document available at the web-site (see text). Nagoya muon detector (MD) consists of two horizontal 6$\times$6 arrays of 1 m$^{2}$ unit detectors, vertically separated by 1.73 m, with an intermediate 5 cm layer of lead to absorb the soft component radiation in the air. Each unit detector has a 1 m $\times$ 1 m plastic scintillator viewed by a photomultiplier tube of 12.7 cm diameter. By counting pulses of the twofold coincidences between a pair of detectors on the upper and lower layers, Nagoya MD records the rate of muons from the corresponding incident direction, as shown in this figure. The multi-directional MD comprises various combinations between the upper and lower detectors. The directional channels named ``30$^{\circ}$W'', ``49$^{\circ}$W'' and ``64$^{\circ}$W'' in this figure correspond to ``W'', ``W2'' and ``W3'' in Table \ref{tab:character}, respectively. The geomagnetic cut-off rigidity ($P_c$) and median primary rigidity ($P_m$) in GV are listed in Table \ref{tab:character}.}
\label{fig:NMD}
\end{figure}

We derive the cosmic ray anisotropy in three dimensions by analyzing the pressure corrected hourly count rates recorded by a muon detector (MD) at Nagoya in Japan during 44 years between 1970 and 2013. Nagoya muon detector is multi-directional and capable of simultaneously monitoring intensities in 17 directional channels of viewing (see figure \ref{fig:NMD}). It has been in operation since 1970 producing a continuous record of cosmic ray intensity over four decades and allowing us to analyze the long term variation of the anisotropy\footnote{Description and data of Nagoya MD are available at \url{http://www.stelab.nagoya-u.ac.jp/ste-www1/div3/muon/muon1.html}.}. Based on our own experience of the long-term observation using plastic scintillators and PMTs similar to Nagoya MD, we estimate that the absolute muon count rate by Nagoya MD has decreased $\sim$10 \% or less in four decades due to the deterioration of detectors. The effect of this deterioration, however, should be negligibly small for the GCR anisotropy analyzed in this paper, because our analysis does not use the absolute count rate but the fractional deviation of the count rate from the daily or monthly mean, as shown later in this section. The median rigidity ($P_m$) of primary GCRs, calculated by utilizing the response function of the atmospheric muons to the primary particles \citep{Mur79}, ranges from 59.4 to 113.7 GV, and the statistical error of hourly count rate ranges between 0.06 \% and 0.28 \% \citep{Oka08}. In this paper, we use 60 GV for the representative $P_m$ of Nagoya MD. The response function has been first calculated for each element in GCRs and then averaged with the weight according to the observed elemental abundance of GCRs \citep{Mur79}.\par
We also derive the anisotropy by analyzing the data recorded during the same period by neutron monitors (NMs), Swarthmore/Newark, Alert/Thule and McMurdo, for each of which $P_m$s is 17 GV \citep{Yas82}. We use the data from Swarthmore/Newark to derive the diurnal anisotropy, while we derive the NS anisotropy from a pair of polar NMs at Thule in Greenland (or Alert in Canada) and McMurdo in Antarctica.\footnote{Description and data of NMs are available at \url{http://neutronm.bartol.udel.edu/} and \url{http://center.stelab.nagoya-u.ac.jp/WDCCR/}.} By comparing anisotropies derived from MD and NMs whose $P_m$ differ by a factor of about 3.5 from each other, we discuss the rigidity dependence of the anisotropy and its long term variation. Table \ref{tab:character} summarizes the cosmic ray data analyzed in this paper.\par
In this section, we describe our analyses of Nagoya MD data, while we derive the anisotropy in free space from NMs in Table \ref{tab:character} following the analyses in Paper I and Paper II. For our analyses of NM data, therefore, readers can refer to those papers.

\subsection{Elimination of short-term events and derivation of the observed diurnal variation}
We begin our analyses with calculating the fractional deviation $\Delta I_j(t)$ of the pressure corrected hourly muon count rate $I_j(t)$ in the $j$-th directional channel of Nagoya MD ($j = 1,2,\ldots,17$) at the universal time $t$ from the 24-hours central moving average $\bar{I}_j^{\rm 24h}(t)$, as
\begin{equation}
   \Delta I_j(t) = \left( I_j(t) - \bar{I}_j^{\rm 24h}(t) \right)/\bar{I}_j^{\rm 24h}(t)\\
   \label{dI}
\end{equation}
where
\begin{equation}
   \bar{I}^{\rm 24h}_j(t) = \frac{1}{24} \sum_{t-12}^{t+11} I_j(t).
\end{equation}
For the following analyses of the diurnal anisotropy, we use $\Delta I_j(t)$ in equation (\ref{dI}) instead of $I_j(t)$ itself to avoid the influence of the gradual intensity variation, such as day-to-day variation, to the diurnal variation. We then check the difference between the maximum and minimum values of $\Delta I_{NM}(t)$ for the McMurdo NM data in every day and exclude the day with the difference exceeding 2.0 \% from further analyses to avoid the influence of large cosmic ray events such as the Forbush decreases. Total 777 days are excluded out of 16,071 days in 44 years between 1970 and 2013 in our analyses of MD and NM data. We confirmed that these excluded days include the majority of cosmic ray events reported so far \citep{Can96, Jor11}. From $\Delta I_j(t)$ in the remaining days, we obtain the monthly mean diurnal distribution, $d_j(t_k)$, of $\Delta I_j(t)$ as a function of the local solar time $t_k (k = 1,2,\ldots,24)$ at the observation site, Nagoya in Japan.\par
We then deduce the diurnal variation of GCR intensity from the Fourier analysis of $d_j(t_k)$ described above, as
\begin{mathletters}
\begin{eqnarray}
   a_{1,j}^{1,\rm obs} = \frac{1}{\pi} \sum_{k=1}^{24} d_j(t_k) \cos(\omega t_k) \label{aobs}\\
   b_{1,j}^{1,\rm obs} = \frac{1}{\pi} \sum_{k=1}^{24} d_j(t_k) \sin(\omega t_k) \label{bobs}
\end{eqnarray}
\end{mathletters}
where $a_{1,j}^{1,\rm obs}$ and $b_{1,j}^{1,\rm obs}$ are the observed harmonic components of the monthly average diurnal variation and $\omega$ is $\pi/12$. In the following subsections, we use $a_{1,j}^{1,\rm obs}$ and $b_{1,j}^{1,\rm obs}$ for deriving the diurnal anisotropy at 60 GV in free space corrected for the geomagnetic effects by taking account of the energy response of each directional channel. We use $a_{1,j}^{1,\rm obs}$ and $b_{1,j}^{1,\rm obs}$ observed by Newark/Swarthmore NM during the same period for deriving the free space diurnal anisotropy at 17 GeV (see Paper II).

\subsection{Correction for the Compton-Getting effect arising from the Earth's orbital motion around the Sun}
The Earth's orbital motion around the Sun causes an apparent anisotropy due to the Compton-Getting (CG) effect \citep{Com35, Cut86, Ame04}. The amplitude and phase of this apparent anisotropy in space are known to be independent of particle's rigidity $p$. Space harmonic components of this anisotropy, $\xi_x^{\rm CG}$ and $\xi_y^{\rm CG}$ in the Geocentric Solar Ecliptic coordinate system (GSE), are given, as
\begin{mathletters}
\begin{eqnarray}
   \xi_x^{\rm CG} = 0,\label{CGx}\\
   \xi_y^{\rm CG} = -(2 + \Gamma) v_{E}/c\label{CGy}
\end{eqnarray}
\end{mathletters}
where $\Gamma$ is the power law index of the energy spectrum of GCRs, $v_E$ is the Earth's velocity and $c$ is the speed of light. We set $\Gamma$ and $v_E$ to be 2.7 and 30 km/s, respectively. Note that we define the anisotropy vector throughout this paper as a vector pointing toward a direction \it{from} \rm which the highest GCR flux is measured; i.e., the anisotropy vector is oppositely directed to the GCR streaming vector.\par
The harmonic components of the diurnal variation expected from this effect for $j$-th directional channel of MD are then given, as
\begin{mathletters}
\begin{eqnarray}
   a_{1,j}^{1,\rm CG} = c_{1,j}^{1,\rm CG} \xi_{x}^{\rm CG(GEO)} + s_{1,j}^{1,\rm CG} \xi_y^{\rm CG(GEO)}\label{a}\\
   b_{1,j}^{1,\rm CG} = -s_{1,j}^{1,\rm CG} \xi_{x}^{\rm CG(GEO)} + c_{1,j}^{1,\rm CG} \xi_y^{\rm CG(GEO)}\label{b}
\end{eqnarray}
\end{mathletters}
where $\xi_x^{\rm CG(GEO)}$ and $\xi_y^{\rm CG(GEO)}$ are the space harmonic components of the CG anisotropy transformed to the Geographic coordinate system (GEO) and $c_{1,j}^{1,\rm CG}$ and $s_{1,j}^{1,\rm CG}$ are so-called the coupling coefficients relating the observed harmonic vector with the space harmonic vector and are calculated \citep{Fuj84}, as
\begin{mathletters}
\begin{eqnarray}
   c_{1,j}^{1,\rm CG} = \frac{1}{\bar{I}_j} \int_{p_{cj}}^{\infty} \int_{\Omega_j} \int_{S_j}
			Y \cdot G^{\rm CG}(p) \cdot P_{1}^{1}(\cos \theta_{\rm or})
			\cdot \cos(\phi_{\rm or} - \phi_{\rm st}) dS d\Omega dp \label{c}\\
   s_{1,j}^{1,\rm CG} = \frac{1}{\bar{I}_j} \int_{p_{cj}}^{\infty} \int_{\Omega_j} \int_{S_j}
			Y \cdot G^{\rm CG}(p) \cdot P_{1}^{1}(\cos \theta_{\rm or})
			\cdot \sin(\phi_{\rm or} - \phi_{\rm st}) dS d\Omega dp. \label{s}
\end{eqnarray}
\end{mathletters}
In equations (\ref{c}) and (\ref{s}), $\bar{I}_j$ is the average count rate in the $j$-th directional channel of muon detector, $Y$ is the response function of the atmospheric muons to primary GCRs and $p_{cj}$ is the cut-off rigidity below which $Y$ is insignificant \citep{Mur79}. The response function $Y$ gives the number of muons produced by primary particles of rigidity $p$ and arriving at $j$-th directional channel with the zenith angle $\theta$ and azimuth angle $\phi$. $P_1^1(x)$ is the semi-normalized spherical function $P_n^m(x)$ with $n = m = 1$ \citep{Cha40}. $S_j$ and $\Omega_j$ are respectively the total area and solid angle of the $j$-th directional channel and $dS$ and $d\Omega$ are those elements. $\phi_{\rm st}$ is the geographic longitude of the detector site and $\theta_{\rm or}$ and $\phi_{\rm or}$ are respectively the geographic co-latitude and longitude defining the asymptotic direction outside the geomagnetic field of primary particles with $p$ which produce muons with the incident direction ($\theta$, $\phi$), as determined using a particle trajectory code \citep{Lin95}. The integrals in equations (\ref{c}) and (\ref{s}) are over all rigidity values for which primary particles produce detectable muons and over all incident directions ($\theta$, $\phi$) for which muon can enter the $j$-th directional channel. In equations (\ref{c}) and (\ref{s}), $G^{\rm CG}$ is the rigidity spectrum of the Compton-Getting anisotropy and independent of $p$, as
\begin{equation}
   G^{\rm CG}(p) = 1.
\end{equation}
\par
Using equations (\ref{a}) and (\ref{b}) as an example, we briefly describe a physical implication of the coupling coefficients. The phase of the CG-anisotropy in space given in equations (\ref{CGx}) and (\ref{CGy}) is 270$^{\circ}$ in the GSE longitude or 06:00 hour in the local solar time. With the coupling coefficients $c_{1,j}^{1,\rm CG}$ and $s_{1,j}^{1,\rm CG}$ in equations (\ref{c}) and (\ref{s}) which are both positive for the vertical channel of Nagoya MD, we get $a_{1,j}^{1,\rm CG}$ and $b_{1,j}^{1,\rm CG}$ in equations (\ref{a}) and (\ref{b}) both positive for this channel, representing the phase of the observed anisotropy shifted to earlier hours from 06:00 hour in space due to the deflection of orbits of positively charged GCRs in the geomagnetic field. In case of $a_{1,j}^{1}$ and $b_{1,j}^{1}$ due to the unknown anisotropy, we can use the coupling coefficients to correct the observed anisotropy for the geomagnetic deflection, by solving equations like (\ref{a}) and (\ref{b}) for the unknown anisotropy ($\xi_{x}, \xi_{y}$) in space.\par
As described below, we correct the observed diurnal variation for the CG effect by subtracting the expected harmonic components $a_{1,j}^{1,\rm CG}$ and $b_{1,j}^{1,\rm CG}$ in equations (\ref{a}) and (\ref{b}) from the observed components $a_{1,j}^{1,\rm obs}$ and $b_{1,j}^{1,\rm obs}$ in equations (\ref{aobs}) and (\ref{bobs}), respectively.

\subsection{Derivation of the three dimensional anisotropy in free space}
The three dimensional (3D) anisotropy of GCR intensity consists of three components, two lying in the ecliptic plane and the third pointing normal to the ecliptic plane. The ecliptic components are observed as the diurnal variation in solar time of GCR intensity recorded with a ground based detector, while the normal component is observed as the north-south (NS) anisotropy. In the following sub sections, we deduce the diurnal anisotropy and the NS anisotropy at 60 GV from Nagoya MD data, while we derive the anisotropy at 17 GV from NM data in Table \ref{tab:character}.

\subsubsection{Modeling harmonic components of the diurnal variation}
The harmonic components $a_{1,j}^{1,\rm obs}$ and $b_{1,j}^{1,\rm obs}$ of the diurnal anisotropy observed by Nagoya MD are expressed in terms of the unknown harmonic components $\xi_x^{\rm GEO}$ and $\xi_y^{\rm GEO}$ representing the diurnal anisotropy in free-space in the Geographic coordinate system (GEO), as
\begin{mathletters}
   \begin{eqnarray}
      a_{1,j}^1 = a_{1,j}^{1,\rm CG} + c_{1,j}^{1} \xi_x^{\rm GEO} + s_{1,j}^{1} \xi_y^{\rm GEO}
		     + a_{\rm com} \label{aBF}\\
      b_{1,j}^1 = b_{1,j}^{1,\rm CG} - s_{1,j}^{1} \xi_x^{\rm GEO} + c_{1,j}^{1} \xi_y^{\rm GEO}
		     + b_{\rm com} \label{bBF}
   \end{eqnarray}
\end{mathletters}
where $c_{1,j}^1$ and $s_{1,j}^1$ are the coupling coefficients given by equations (\ref{c}) and (\ref{s}) with $G^{\rm CG}(p)$ replaced with $G(p)$ for the unknown rigidity spectrum of the diurnal anisotropy. In equations (\ref{aBF}) and (\ref{bBF}), $a_{\rm com}$ and $b_{\rm com}$ are harmonic components of the diurnal variation arising from the atmospheric temperature effect on muon intensity which is assumed in this paper to be common for all directional channels as the first-order approximation. For $G(p)$, we assume in this paper, 
\begin{eqnarray}
   G(p) = 1 \;{\rm for}\; p \le P_u  \nonumber\\
        = 0 \;{\rm for}\; p > P_u  \label{G}
\end{eqnarray}
where $P_u$ is the upper limiting rigidity of the anisotropy and set to be constant at 100 GV. This spectrum is used in Paper II for the analysis of NM data and we use the same spectrum for MD data as well. Results derived with different $P_u$ are shown and discussed in Appendix B.

\subsubsection{Deriving the diurnal anisotropy in free space}
\begin{figure}
\plotone{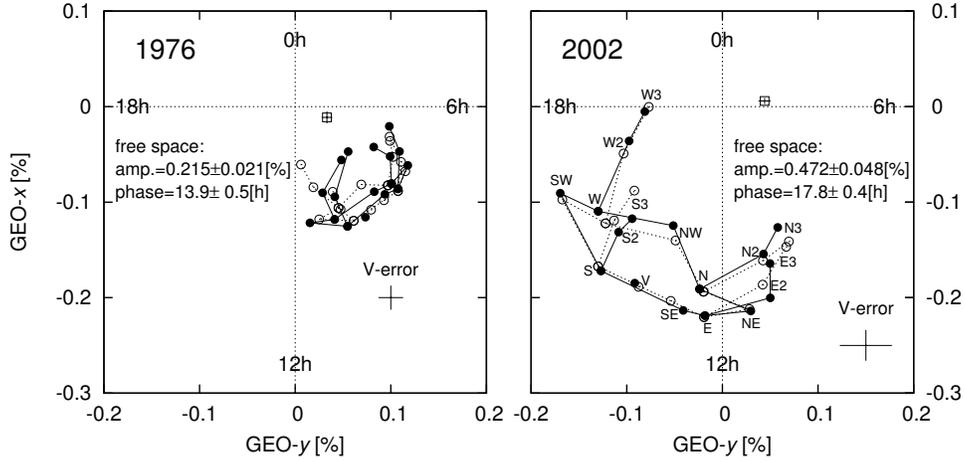}
\caption{Yearly mean harmonic dials of the diurnal anisotropy observed by the Nagoya multi-directional muon detector in 1976 (left) around $A>0$ solar activity minimum and in 2002 (right) around $A<0$ solar activity maximum. Solid circles display the harmonic vector ($a_{1,j}^{1,\rm obs}$, $b_{1,j}^{1,\rm obs}$) observed by the $j$-th directional channel with $a_{1,j}^{1,\rm obs}$ and $b_{1,j}^{1,\rm obs}$ plotted on the vertical (GEO-$x$) and horizontal (GEO-$y$) axes, respectively, while open circles display the best-fit vectors. The phases of the diurnal anisotropy with $x>0$ and $y=0$, $x=0$ and $y>0$, $x<0$ and $y=0$ and $x=0$ and $y<0$ are 00:00, 06:00, 12:00, 18:00 hours in the local solar time , respectively. To demonstrate the relative configuration of the observed (best-fit) harmonic vectors in 17 directional channels, the head of each vector is connected with each other by solid (dotted) thin lines (see directional channels indicated in the right panel). An open square with an error cross in each panel displays the common vector representing the atmospheric temperature effect. Amplitude and phase of the best-fit harmonic vector in free space are indicated in each panel. For reference, the cross in the bottom-right corner in each panel represents errors of $a_{1,j}^{1,\rm obs}$ and $b_{1,j}^{1,\rm obs}$ in vertical (V) channel, deduced from the dispersion of monthly values.}
\label{fig:harm}
\end{figure}

We deduce the best-fit parameters $\xi_x^{\rm GEO}$, $\xi_y^{\rm GEO}$, $a_{\rm com}$ and $b_{\rm com}$ in equations (\ref{aBF}) and (\ref{bBF}) that minimize the residual $S$, defined as
\begin{equation}
   S = \sum_{j=1}^{17} \left\{ \left(a_{1,j}^{1,\rm obs} - a_{1,j}^{1}\right)^2/\sigma_{a,j}^2
	 + \left(b_{1,j}^{1,\rm obs} - b_{1,j}^{1}\right)^2/\sigma_{b,j}^2 \right\}
   \label{chi2}
\end{equation}
where $\sigma_{a,j}$ and $\sigma_{b,j}$ are errors of $a_{1,j}^{1,\rm obs}$ and $b_{1,j}^{1,\rm obs}$, respectively, and deduced from the dispersion of $\Delta I_j(t)$ used for calculating the monthly mean $d_j(t_k)$ at the local time $t_k$ in equations (\ref{aobs}) and (\ref{bobs}). We perform this calculation for every month and calculate yearly mean values and errors of $\xi_x^{\rm GEO}$, $\xi_y^{\rm GEO}$, $a_{\rm com}$ and $b_{\rm com}$ from means and dispersions of 12 monthly values, respectively. Figure \ref{fig:harm} displays sample comparisons between the best-fit and the observed yearly mean harmonic vectors for Nagoya MD in 2002 and 1976 when the solar activity were close to the maximum and minimum, respectively. It is clear that the amplitude of the derived space harmonic vector indicated in each panel is significantly larger in 2002 than that in 1976 causing an ``expansion'' of the pattern drawn by lines connecting heads of harmonic vectors observed by 17 directional channels during the solar maximum period. It is also clear that the phase of the derived space harmonic vector is about 4 hours earlier in 1976 than in 2002, due to the 22-year variation of the diurnal anisotropy.

\subsubsection{Identification of IMF sector and solar dipole magnetic field polarities}
In order to calculate the diurnal anisotropy in each IMF sector, we identify the sector polarity ({\it toward} or {\it away}) of each day referring to the polarity of the Stanford Mean Magnetic Field of the Sun (WSO web-site at \url{http://wso.stanford.edu/}) with the date shifted 5 days later for a rough correction for the solar wind transit time between the Sun and the Earth. For the period prior to 1975 when the data are not available on the WSO web-site, we identify the polarity by the IMF data in the National Space Science Data Center's ``omnitape'' \citep{Kin05} following the analysis by Paper II. Because of serious gaps in the ``omnitape'' data particularly in 1980's and 1990's, we give it up to use the ``omnitape'' IMF data for an entire period of the present analysis. By analyzing a period when both the Stanford Mean Magnetic Field and the ``omnitape'' data are available, we confirmed that the daily sector polarities identified by these two methods are quite consistent with each other, giving the essentially same results from our cosmic ray data analyses.\par
We then calculate the average diurnal distribution, $d_j^T(t_k)$ ($d_j^A(t_k)$), for {\it toward} ({\it away}) days in every month. By using $d_j^T(t_k)$ ($d_j^A(t_k)$) for $d_j(t_k)$ in equations (\ref{aobs}) and (\ref{bobs}) and for the best-fit calculation described above, we obtain $\xi_x^{{\rm GEO}(T)}$, $\xi_y^{{\rm GEO}(T)}$, $a_{\rm com}^T$ and $b_{\rm com}^T$ ($\xi_x^{{\rm GEO}(A)}$, $\xi_y^{{\rm GEO}(A)}$, $a_{\rm com}^A$ and $b_{\rm com}^A$) in {\it toward} ({\it away}) sector in every month. Monthly mean parameters are then calculated by taking mean of {\it toward} and {\it away} values.\par
For the following discussions of yearly mean parameter, we also assign the polarity of the large-scale solar magnetic field for each year referring to the ``Solar Polar Field Strength'' available at the WSO web-site where the average polar field strength is given in every Carrington Rotation. We assign the polarity of a year as $A>0$ ($A<0$) when the average polar field in the year is positive pointing away from the Sun in the northern (southern) hemisphere. We regard a year as a period of the polarity reversal in progress when the year contains Carrington Rotations with the polar field pointing {\it away} or {\it toward} in both hemispheres. For a period prior to 1975 when the WSO data are unavailable, we follow the assignment by Paper I. The polarity of each year assigned by us is indicated in Table \ref{tab:xiMD} in Appendix A.\par
Figure \ref{fig:diurnal} displays temporal variations of the amplitude (upper panel) and phase (lower panel) of the yearly mean harmonic vector in free-space. Clearly seen in this figure is the phase in the lower panel showing a prominent 22-year variation, with minima occurring in 1976 and 1997 around $A>0$ solar minima. This phase variation is about $\sim$2 hours in NM data (open circles), while it is almost double ($\sim$4 hours) in MD data (solid circles). The amplitude of the diurnal anisotropy in the upper panel is smaller (larger) around the solar minimum (maximum) period in both the NM and MD data. Table \ref{tab:xiMD} in Appendix A lists numerical data of best-fit parameters obtained for each year. As shown in figure \ref{fig:common} in Appendix A, the mean amplitude of the common vector $(a_{\rm com},b_{\rm com})$ in equations (\ref{aBF}) and (\ref{bBF}), which is introduced to represent the atmospheric temperature effect, is small ($0.039\pm0.002$ \%), while the phase is almost stable around $\sim$06:00 local time in an agreement with the average temperature effect reported from muon observations \citep[e.g.][]{Mun97}. It is also seen in figure \ref{fig:common} that the common vector shows no notable long-term variations in correlation with the solar activity- or magnetic-cycle.

\subsubsection{Derivation of the north-south anisotropy}
We derive the north-south (NS) anisotropy perpendicular to the ecliptic plane at 60 GV from the Nagoya GG-component (see Paper II for the derivation of the north-south anisotropy from NM data). The GG-component is a difference combination between intensities recorded in the north- and south-viewing channels designed to represent the NS anisotropy free from the atmospheric temperature effect \citep{Nag72,Mor79}. The GG-component is defined, as
\begin{equation}
   GG(t) = \left\{r_{\rm N2}(t) - r_{\rm S2}(t)\right\} + \left\{r_{\rm N2}(t) - r_{\rm E2}(t)\right\}
\end{equation}
where $r_{XX}(t)$ is the percent deviation of the pressure-corrected muon rate $I_{XX}(t)$ in the directional channel XX(= N2, S2, E2) from the monthly mean. We calculate $GG^T$ and $GG^A$ by averaging $GG(t)$ over {\it toward} and {\it away} days, respectively, according to the IMF sector polarity in every month and calculate the difference, $\Delta GG$, as
\begin{equation}
   \Delta GG = (GG^T - GG^A)/2.
   \label{dGG}
\end{equation}
The NS anisotropy $\xi_z^{{\rm GEO}(T)}$ in space in {\it toward} sector is calculated in every month from $\Delta GG$, as
\begin{equation}
   \xi_z^{{\rm GEO}(T)} = \Delta GG/\left( 2 c_{1,{\rm N2}}^0 - c_{1,{\rm S2}}^0 - c_{1,{\rm E2}}^0 \right)
\end{equation}
where $c_{1,XX}^0$ is the coupling coefficient for the directional channel XX given, as
\begin{equation}
   c_{1,XX}^0 = \frac{1}{\bar{I}_{XX}} \int_{p_{cXX}}^{\infty} \int_{\Omega_{XX}} \int_{S_{XX}}
			Y \cdot G(p) \cdot P_{1}^{0}(\cos \theta_{\rm or}) dS d\Omega dp
\end{equation}
with the rigidity spectrum $G(p)$ in (\ref{G}). We deduce $\xi_z^{\rm GEO}$ from the difference between GG-components in {\it toward} and {\it away} days ($\Delta GG$) in equation (\ref{dGG}) because of the assumption that the anisotropy vector, when averaged over one month, is symmetric with respect to the heliospheric current sheet (HCS) and the NS anisotropy lies in an opposite direction with the same magnitude above and below the HCS, as
\begin{equation}
   \xi_z^{{\rm GEO}(A)} = -\xi_z^{{\rm GEO}(T)}.
\end{equation}

\subsection{Derivation of modulation parameters}
\subsubsection{Anisotropy components in the solar wind frame}
\begin{figure}
\plotone{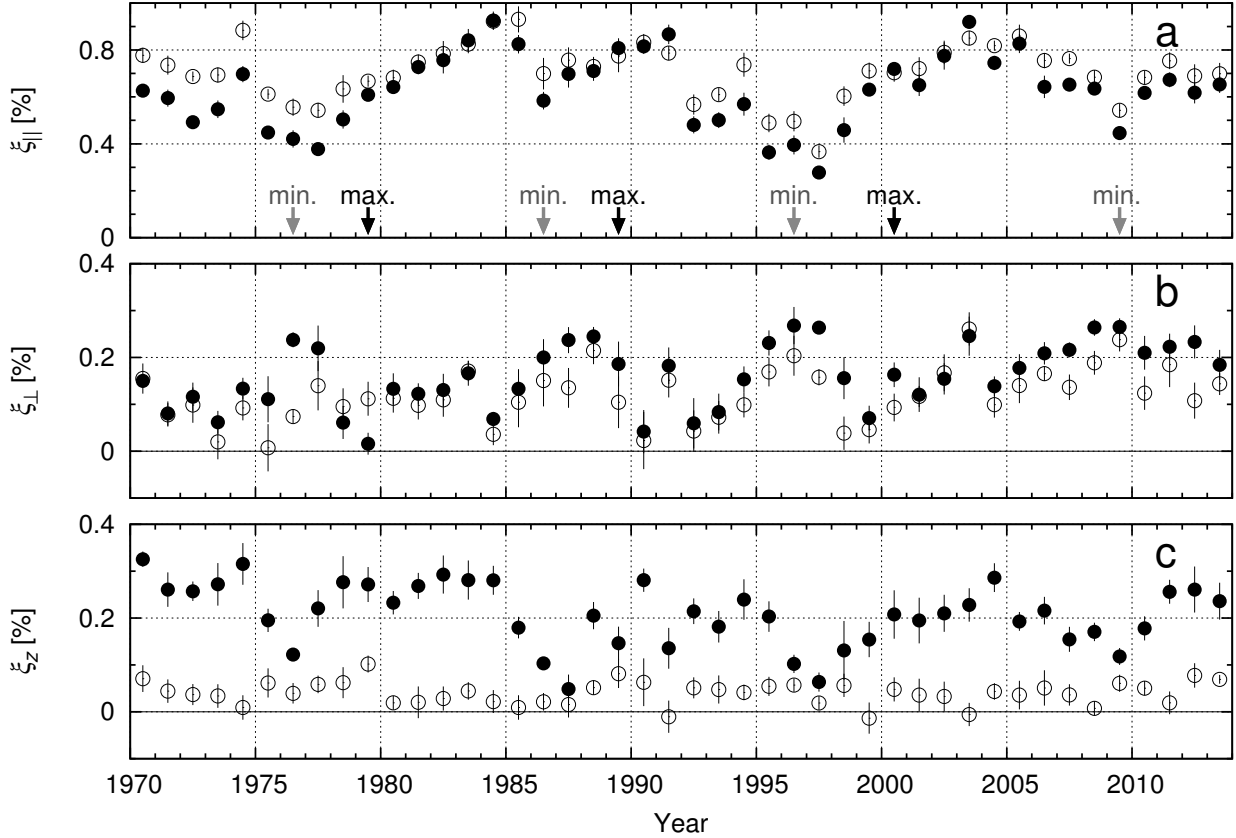}
\caption{Three components of the anisotropy in the solar wind frame. Each panel from top to bottom displays the yearly mean $\xi_{\|}$, $\xi_{\bot}$ and $\xi_{z}$ in \% as a function of year. Solid circles display the anisotropy components derived from MD data at 60 GV, while open circles show the anisotropy derived from NM data at 17 GV (see Table \ref{tab:xiMD} in Appendix A for numerical data from MD). Yearly mean and error are deduced from the mean and dispersion of monthly values, respectively. The solar maximum and minimum periods are indicated by black and gray arrows on the horizontal axis of the top panel, respectively.}
\label{fig:IMFcomp}
\end{figure}

Three components ($\xi_x^{{\rm GEO}(T/A)}$, $\xi_y^{{\rm GEO}(T/A)}$, $\xi_z^{{\rm GEO}(T/A)}$) of the space anisotropy vector obtained above are first converted to components ($\xi_x^{(T/A)}$, $\xi_y^{(T/A)}$, $\xi_z^{(T/A)}$) in the Geocentric Solar Ecliptic coordinate system (GSE) and then transformed to the solar wind frame for deriving the modulation parameters. We obtain the anisotropy components ($\xi_x^{{\rm SW}}$, $\xi_y^{{\rm SW}}$, $\xi_z^{{\rm SW}}$) in the solar wind frame by subtracting the contribution from the solar wind convection, as
\begin{mathletters}
\begin{eqnarray}
   \xi_x^{{\rm SW}(T/A)} = \xi_x^{(T/A)} - (2 + \Gamma) V_{\rm SW}^{(T/A)}/c\\
   \xi_y^{{\rm SW}(T/A)} = \xi_y^{(T/A)}\\
   \xi_z^{{\rm SW}(T/A)} = \xi_z^{(T/A)}
\end{eqnarray}
\end{mathletters}
where $V_{\rm SW}^{(T/A)}$ is the radial component of the solar wind velocity in the omnitape data \citep{Kin05}. We then calculate parallel and perpendicular components of the anisotropy, as
\begin{mathletters}
\begin{eqnarray}
   \xi_{\|}^{(T/A)} = \xi_x^{{\rm SW}(T/A)} b_x^{(T/A)} + \xi_y^{{\rm SW}(T/A)} b_y^{(T/A)} \label{xpar}\\
   \xi_{\bot}^{(T/A)} = -\xi_x^{{\rm SW}(T/A)} b_y^{(T/A)} + \xi_y^{{\rm SW}(T/A)} b_x^{(T/A)} \label{xper}
\end{eqnarray}
\end{mathletters}
where $b_x^{(T/A)}$ and $b_y^{(T/A)}$ are GSE components of a unit vector pointing away from the Sun along the IMF and calculated from the mean IMF in the omnitape data. Note that positive $\xi_{\|}^{(T/A)}$ and $\xi_{\bot}^{(T/A)}$ correspond to the GCR streaming inward to the inner heliosphere parallel and perpendicular to IMF, respectively. We finally obtain monthly average components of the anisotropy in the solar wind frame, as
\begin{mathletters}
\begin{eqnarray}
   \xi_{\|} = \left(\xi_{\|}^{(T)} + \xi_{\|}^{(A)}\right)/2\\
   \xi_{\bot} = \left(\xi_{\bot}^{(T)} + \xi_{\bot}^{(A)}\right)/2\\
   \xi_{z} = \left(\xi_{z}^{(T)} - \xi_{z}^{(A)}\right)/2.
\end{eqnarray}
\end{mathletters}
This definition of $\xi_{z}$ is again from the assumption of the symmetry above and below the HCS. Note that the positive $\xi_{z}$ corresponds to the southward GCR streaming perpendicular to the ecliptic plane in the {\it toward} IMF sector. We perform calculations of $\xi_{\|}$, $\xi_{\bot}$, $\xi_{z}$ described above in every month and deduce the yearly mean value and its error of each anisotropy component from the mean and dispersion of 12 monthly values, respectively. Figure \ref{fig:IMFcomp} shows $\xi_{\|}$, $\xi_{\bot}$, $\xi_{z}$ each as a function of year. It is seen that three components of the anisotropy derived from MD data (solid circles) are all positive throughout the entire period in this figure. A clear 22-year variation seen in $\xi_{\|}$ in figure \ref{fig:IMFcomp}a indicates that this component anisotropy is responsible for the phase variation in figure \ref{fig:diurnal} as discovered in Paper I and Paper II. No such clear signature of 22-year variation is seen in either $\xi_{\bot}$ or $\xi_{z}$ displayed in figures \ref{fig:IMFcomp}b and \ref{fig:IMFcomp}c.\par
There is a close correlation between the variation of the $\xi_{\|}$ values obtained for NMs at 17 GV and for MD at 60 GV (open and solid circles in figure \ref{fig:IMFcomp}a, respectively), indicating a weak rigidity dependence of this anisotropy component. A scatter plot of $\xi_{\|}$ for NMs and that for MD on the $x$ and $y$ axes, respectively, yields a correlation coefficient $r$=0.92 and a slope (ratio) of $y/x=\beta=$0.89$\pm$0.05, which suggests that $\xi_{\|}$ remains nearly constant despite the factor of 3.5 difference between the rigidity ranges monitored by NM and MD. On the other hand, we find $\beta=$0.77$\pm$0.07 for $A>0$ which is significantly smaller than the value of $\beta=$0.94$\pm$0.005 found for the $A<0$ epochs, showing that the rigidity spectrum of $\xi_{\|}$ is softer in the $A>0$ epochs. We also see a remarkable correlation between $\xi_{\bot}$ for NMs and that for MD with $r=0.75$, while the $\beta$ values turn out to be 1.65$\pm$0.35 (1.26$\pm$0.14) in $A>0$ ($A<0$) epochs, which indicates that $\xi_{\bot}$ increases with increasing $P_m$. The most significant difference between NM and MD data is seen in the magnitude of $\xi_z$ shown in figure \ref{fig:IMFcomp}c. For this component, we obtain $\beta=$4.45$\pm$0.61 (6.08$\pm$0.96) for the $A>0$ ($A<0$) epochs, which implies that $\xi_z$ increases with increasing rigidity. The correlation between NM and MD data is, however, quite poor ($r=$0.20) for this component. These features appearing in figure \ref{fig:IMFcomp} are qualitatively consistent with $\xi_{\bot}$ and $\xi_z$ increasing with rigidity. The ratios $\beta$ for the three anisotropy components are listed in the column of ``$P_u$=100 GV'' in Table \ref{tab:beta} in Appendix B.\par
We cannot derive any quantitative conclusions about the rigidity spectrum of the anisotropy from the present analysis which assumes a priori a flat spectrum with the upper limiting rigidity $P_u$ fixed at 100 GV as denoted in equation (\ref{G}). Each value of ratios ($\beta$s) described above, for instance, changes for different value of $P_u$. The rigidity dependences of $\xi_{\|}$, $\xi_{\bot}$ and $\xi_{z}$ relative to each other, however, remain unchanged even for different value of $P_u$ (see Appendix B). We will discuss the physical origin of these rigidity dependences in the next subsection.

\subsubsection{Modulation parameters}
Three components ($\xi_{\|}$, $\xi_{\bot}$, $\xi_{z}$) of the anisotropy vector in the solar wind frame obtained above are related to the modulation parameters, i.e. the spatial gradients of GCR density and mean free paths of the pitch angle scattering of GCRs in the turbulent magnetic field, as
\begin{mathletters}
\begin{eqnarray}
   \xi_{\|} = \lambda_{\|} G_r \cos \psi \label{xparG}\\
   \xi_{\bot} = \lambda_{\bot} G_r \sin \psi - R_L G_z \label{xperG}\\
   \xi_{z} = R_L G_r \sin \psi + \lambda_{\bot} G_z \label{xzG}
\end{eqnarray}
\end{mathletters}
where $\lambda_{\|}$ and $\lambda_{\bot}$ are mean free paths of the pitch angle scattering parallel and perpendicular to the IMF, respectively, $R_L$ is the Larmor radius of GCRs in the IMF and $\psi$ is the IMF spiral angle between the radial direction and a unit vector ${\bf b}$ in (\ref{xpar}) and (\ref{xper}) pointing away from the Sun along the IMF. $G_r$ and $G_z$ are the radial and latitudinal components of the fractional density gradient vector defined, as
\begin{equation}
   {\bf G} = \nabla U/U\\
\end{equation}
where $U$ is the GCR density (or omnidirectional intensity) given as a function of the position in the heliosphere, time and GCR rigidity. We assume that the longitudinal gradient is zero in our analyses based on the anisotropy averaged over one month which is longer than the solar rotation period. Note that $G_z$ represents the latitudinal density gradient in {\it toward} sector, being positive when $U$ increases with increasing latitude, and changes its sign in {\it away} sector due to the assumed symmetry above and below the HCS. The bidirectional latitudinal density gradient $G_{|z|}$, which is defined to be positive (negative) when $U$ increases away from (toward) the HCS, is given by $G_z$, as
\begin{equation}
   G_{|z|} = -{\rm sgn}(A) G_z\\
   \label{A}
\end{equation}
where $A$ represents the polarity of the solar dipole magnetic moment and
\begin{eqnarray}
   {\rm sgn}(A) &=& +1, \;{\rm for}\; A>0 \;{\rm epoch}, \nonumber\\
                &=& -1, \;{\rm for}\; A<0 \;{\rm epoch}. \nonumber
\end{eqnarray}\par
Equations (\ref{xparG})-(\ref{xzG}) include four unknown modulation parameters, $\lambda_{\|}$, $\lambda_{\bot}$, $G_r$ and $G_z$, while we have only three components ($\xi_{\|}$, $\xi_{\bot}$, $\xi_{z}$) of the observed anisotropy. We therefore assume in this paper
\begin{equation}
   \lambda_{\bot}/\lambda_{\|} = \alpha = 0.01
\end{equation}
and derive three remaining parameters, $\lambda_{\|}$, $G_r$ and $G_z$. Papers I and II also adopted the same constant value of $\alpha$ based on empirical determinations of $\lambda_{\|}\approx 0.5 AU$ by \citet{Bie83} and $\lambda_{\bot}\approx 0.007 AU$ by \citet{Pal82} for $\sim$10 GV GCRs. From (\ref{xparG}), we get
\begin{equation}
   G_r = \xi_{\|}/\left( \lambda_{\|} \cos \psi \right).
\end{equation}
Introducing this into (\ref{xperG}), we get
\begin{equation}
   G_z = \left( \alpha \xi_{\|} \tan \psi - \xi_{\bot} \right)/R_L.
   \label{Gz}
\end{equation}
From (\ref{xparG}), on the other hand, we also get
\begin{equation}
   \lambda_{\|} = \xi_{\|}/\left( G_r \cos \psi \right).
   \label{lpar}
\end{equation}
Introducing (\ref{Gz}) and (\ref{lpar}) into (\ref{xzG}), we get a quadratic equation for $G_r$, as
\begin{equation}
   R_L \sin \psi G_r^2 - \xi_z G_r - \alpha \xi_{\|} \left( \xi_{\bot} - \alpha \xi_{\|} \tan \psi \right)/\left( R_L \cos \psi \right) = 0
\end{equation}
which has a solution for positive $G_r$, as
\begin{equation}
   G_r = \left\{ \xi_z + \sqrt{\xi_z^2 + 4 \alpha \xi_{\|} \tan \psi \left( \xi_{\bot} - \alpha \xi_{\|} \tan \psi \right)} \right\}/\left( 2 R_L \sin \psi \right).
   \label{Gr}
\end{equation}
We first calculate $G_{|z|}$ and $G_r$ from equations (\ref{Gz}) and (\ref{Gr}), respectively, for every month. We then deduce the yearly mean and its error of each parameter from the mean and dispersion of 12 monthly values, respectively. We do not use equation (\ref{lpar}) for calculating monthly value of $\lambda_{\|}$, because $G_r$, particularly derived from NM data, becomes close to zero in some months resulting in an extremely large $\lambda_{\|}$ and large error of yearly mean $\lambda_{\|}$. We instead derive yearly mean $\lambda_{\|}$ from yearly mean $G_r$ and $\cos \psi$ in equation (\ref{lpar}) and deduce the error by propagating from errors of yearly mean $G_r$ and $\cos \psi$. For $R_L$ for MD and NM data, we use gyro-radii of 60 GV and 17 GV GCRs, respectively, in the monthly mean IMF with the magnitude calculated from the omnitape data.\par
\begin{figure}
\plotone{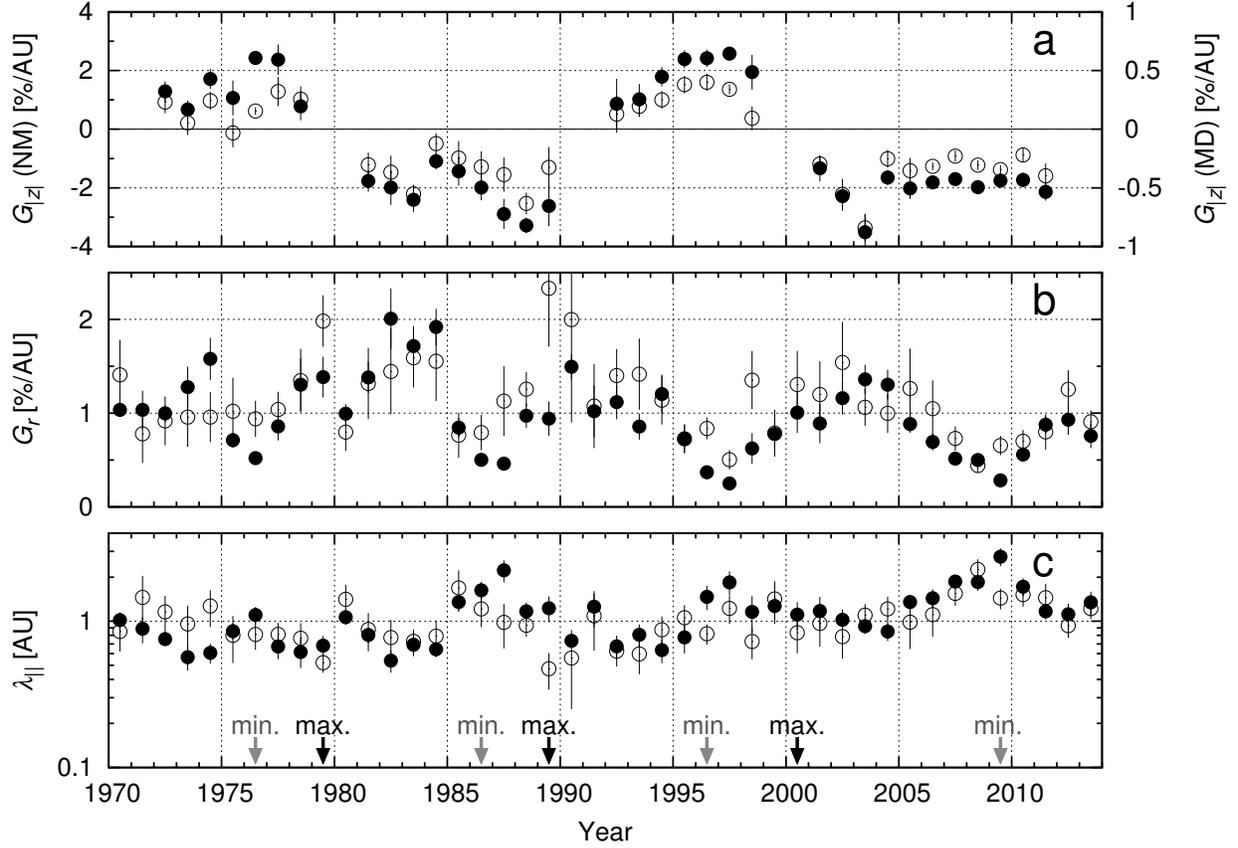}
\caption{Modulation parameters derived from the three dimensional anisotropy. Each panel from top to bottom displays the yearly mean $G_{|z|}$, $G_r$ and $\lambda_{\|}$, each as a function of year. Solid circles display parameters derived from MD data at 60 GV, while open circles show parameters derived from NM data at 17 GV (see Table \ref{tab:xiMD} in Appendix A for numerical data from MD). Note that the bidirectional latitudinal density gradient ($G_{|z|}$) in the top panel is defined to be positive (negative) when the spatial distribution of GCR density has a local minimum (maximum) on the HCS. $G_{|z|}$ and $G_r$ in the top and middle panels are plotted on the vertical axis in linear scales, while $\lambda_{\|}$ in the bottom panel is plotted in a logarithmic scale. Yearly mean and error are deduced from the mean and dispersion of monthly values, respectively. Because of the definition in equation (\ref{A}), $G_{|z|}$ is not available in a year when the polarity reversal is in progress. The solar maximum and minimum periods are indicated by black and gray arrows on the horizontal axis of the bottom panel, respectively.}
\label{fig:ModParam}
\end{figure}

Figure \ref{fig:ModParam} shows the temporal variations of the calculated modulation parameters, $G_{|z|}$, $G_r$ and $\lambda_{\|}$. Clearly seen in figure \ref{fig:ModParam}a is that the bidirectional latitudinal density gradient ($G_{|z|}$) is positive (indicating the local minimum of density on the HCS) in $A>0$ epoch, while it is negative (indicating the local maximum of density on the HCS) in $A<0$ epoch, in accord with the drift model prediction \citep{Kot83}. There is no clear signature of an 11-year variation in $G_{|z|}$. The 22-year variation of $G_{|z|}$ appears cleaner and statistically more significant with relatively smaller errors in MD data than in NM data. The mean $G_{|z|}$ derived from MD (NM) data is 0.42$\pm$0.05 (0.86$\pm$0.14) \%/AU in $A>0$, while it is -0.52$\pm$0.04 (-1.47$\pm$0.15) \%/AU, indicating that the magnitude of $G_{|z|}$ is larger in $A<0$ than in $A>0$ in both MD and NM data.\par
The radial density gradient ($G_r$) in figure \ref{fig:ModParam}b, on the other hand, varies with $\sim$11-year solar activity cycle with maxima (minima) in solar maximum (minimum) periods \citep[Papers I and II;][]{Bie86}, but there is no significant difference seen between mean $G_r$s in $A>0$ and $A<0$ epochs. The mean $G_r$ deduced from MD (NM) data is 0.89$\pm$0.11 (1.04$\pm$0.08) \%/AU in $A>0$ epoch, while it is 0.99$\pm$0.12 (1.13$\pm$0.10) \%/AU in $A<0$ epoch. It is noted that we find a poor correlation between temporal variations of $G_{|z|}$ and $G_r$ in both NM and muon data.\par
The parallel mean free path ($\lambda_{\|}$) in figure \ref{fig:ModParam}c also changes with the solar activity cycle with minima (maxima) in solar maximum (minimum) periods. The mean $\lambda_{\|}$ deduced from MD (NM) data is 0.90$\pm$0.10 (0.89$\pm$0.06) AU in $A>0$, while it is 1.32$\pm$0.13 (1.14$\pm$0.10) AU in $A<0$. This indicates that the mean $\lambda_{\|}$ is systematically larger in $A<0$ than in $A>0$ at 2 or 3 sigma level. It is also interesting that $\lambda_{\|}$s in NM and MD data appear like persistently increasing toward maxima in 2008 and 2009 during the last three solar activity cycles, while $G_r$s look like decreasing. The parallel mean free path ($\lambda_{\|}$) deduced from NM data (open circles) shows peaks in 1985 and 2008 in $A<0$ solar minimum epochs, while it shows smaller peaks in 1974 and 1997 in $A>0$ solar minimum epochs. This is qualitatively consistent with results reported in Paper I. In $\lambda_{\|}$ deduced from MD data (solid circles), on the other hand, the 11-year variation is more prominent with maxima in every solar minimum in 1976, 1987, 1997 and 2009, but no clear 22-year variation is visible in this figure. We will discuss long-term variations of $G_r$ and $\lambda_{\|}$ in more detail in the next section.\par
\begin{figure}
\plotone{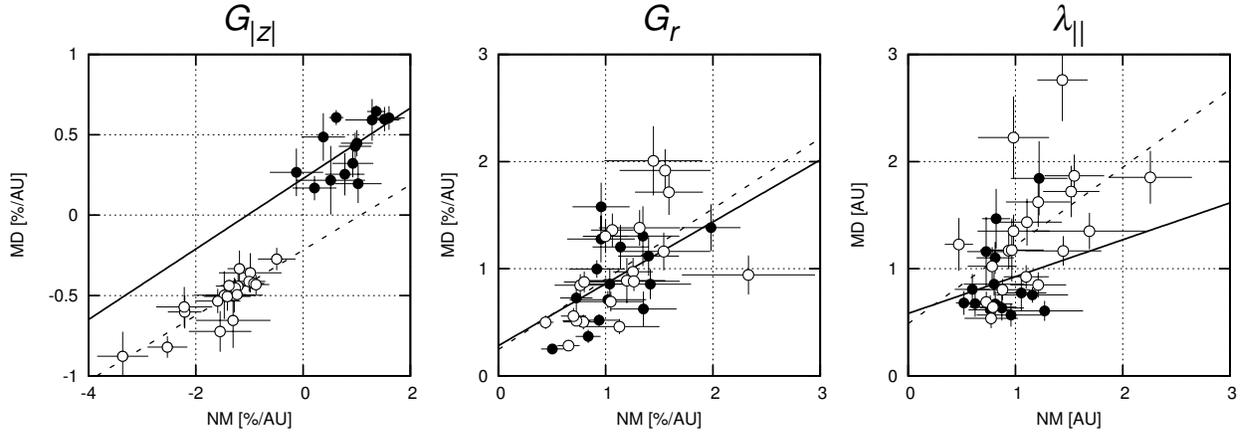}
\caption{Correlation plots between modulation parameters derived from NM data at 17 GV and MD data at 60 GV. The left, middle and right panels show correlations of $G_{|z|}$, $G_r$ and $\lambda_{\|}$, respectively. Each panel displays the parameter in figure \ref{fig:ModParam} derived from MD data at 60 GV on the vertical axis as a function of the parameter derived from NM data at 17 GV in the same year on the horizontal axis. Solid and open circles in each panel display parameters in $A>0$ and $A<0$ epochs, respectively. Data points in years when the polarity reversal is in progress are omitted in this figure.}
\label{fig:NM-Nag}
\end{figure}

We now discuss the rigidity dependence of each modulation parameter. Figure \ref{fig:NM-Nag} shows the correlation between the parameters derived from NM data at 17 GV and from MD data at 60 GV. In $A>0$ ($A<0$) epoch, $G_{|z|}$ from NM and MD data in the left panel shows a good correlation with $r$ of 0.63 (0.86), while the mean ratio ($\beta = y/x$) of $G_{|z|}$ from MD data to that from NM data is 0.48$\pm$0.10 (0.35$\pm$0.05) in $A>0$ ($A<0$) epoch indicating that $G_{|z|}$ decreases with increasing $P_m$. Also similar but weaker correlations are seen in $G_r$ and $\lambda_{\|}$ in the middle and right panels with the average $r$ of 0.53 (0.58) and 0.21 (0.54), respectively, while the mean $\beta$s of $G_r$ and $\lambda_{\|}$ are 0.85$\pm$0.12 (0.87$\pm$0.13) and 1.00$\pm$0.13 (1.16$\pm$0.15), respectively, indicating that these parameters are almost independent of $P_m$. Note that $\beta$ of $G_{|z|}$ is significantly smaller than $\beta$ of $G_{r}$ indicating the softer rigidity dependence of $G_{|z|}$ than that of $G_r$, when $P_u$ is fixed at 100 GV. The ratio $\beta$s derived from different $P_u$ are listed in Table \ref{tab:beta} in Appendix B.\par
We finally discuss the physical origin of the rigidity dependence of each anisotropy component presented in the preceding subsection. As expressed in equations (\ref{xparG})-(\ref{xzG}), $\xi_{\bot}$ and $\xi_z$ include contributions from the drift (i.e. the diamagnetic drift) added to the perpendicular diffusion, while $\xi_{\|}$ results solely from the parallel diffusion. By using $G_r$, $G_{|z|}$ and $\lambda_{\|}$ with an assumption of $\lambda_{\bot} = \alpha \lambda_{\|} = 0.01\lambda_{\|}$, we calculate individual contributions from the diffusion and drift to each of $\xi_{\bot}$ and $\xi_z$. We find that the mean diffusion contribution ($\lambda_{\bot} G_r \sin \psi$) to $\xi_{\bot}$ is significantly smaller than the mean drift contribution ($-R_L G_z$) in both NM and MD data, hence $\xi_{\bot}$ is mainly arising from the drift effect. The mean ratio of $|\lambda_{\bot} G_r \sin \psi|$ to $|-R_L G_z|$ contributing to $\xi_{\bot}$ is 0.08$\pm$0.02 in NM data, while the ratio is 0.07$\pm$0.02 in MD data indicating that the mean contribution from the diffusion to $\xi_{\bot}$ is less than 10 \% in both NM and MD data independent of $P_m$. The mean ratio of the diffusion ($|\lambda_{\bot} G_z|$) to the drift ($|R_L G_r \sin \psi|$) contributing to $\xi_z$ is also small as 0.03$\pm$0.01 in MD data. The ratio in NM data, on the other hand, is 0.19$\pm$0.03 and significantly larger than the ratio in MD data, indicating that the relative contribution of the diffusion to $\xi_z$ increases with decreasing $P_m$. This is due to the rigidity dependence of $G_{|z|}$, which is softer than that of $G_r$ as discussed above. Since there is only a poor correlation between temporal variations of $G_{|z|}$ and $G_r$ in figure \ref{fig:ModParam}, this may explain the poor correlation between $\xi_z$s by NM and MD data which is shown in the bottom panel of figure \ref{fig:IMFcomp} and discussed in the preceding subsection.

\section{Summary and Discussions}
We examined the energy dependence of the long-term variations of the 3D anisotropy of GCR intensity by analyzing the data recorded in 1970-2013 by NMs (Swarthmore/Newark, Alert/Thule and McMurdo) which have median responses to $\sim$17 GV primary GCRs and the Nagoya MD which has the median response to $\sim$60 GV GCRs. The derived free-space harmonic vector of the diurnal anisotropy changes its phase to earlier hours in $A>0$ solar minima from the $\sim$18:00 local time known as the phase of the ``corotation'' anisotropy, while the amplitude changes in 11-year cycle decreasing to a small value in years around every solar minimum. We note that the magnitude of the phase change is significantly larger in MD data than in NM data indicating a marked rigidity dependence of the phase change. A clear 22-year variation is seen in the parallel component ($\xi_{\|}$) of the anisotropy confirming the conclusion of Paper II that $\xi_{\|}$ is primarily responsible for the phase change. The north-south anisotropy ($\xi_z$) derived from the GG-component of Nagoya MD also shows an 11-year cycle with minima in years around every solar minimum.\par
The ecliptic anisotropy components ($\xi_{\|}$ and $\xi_{\bot}$) derived from NM and MD data vary in a close correlation with each other, while no such correlation is seen in the variation of $\xi_z$. The mean ratio between $\xi_{\|}$s in MD and that in NM data is roughly consistent with a rigidity independent spectrum, while the rigidity spectrum of $\xi_{\|}$ is systematically softer in $A>0$ than in $A<0$. 
On the other hand, $\xi_{\bot}$ and $\xi_z$ derived from MD data are significantly larger than those from NM data, indicating that these components increase with $P_m$. According to equations (\ref{xparG})-(\ref{xzG}), $\xi_{\bot}$ and $\xi_z$ include contributions from the gyration of particles (connected to diamagnetic drift) added to perpendicular diffusion, while $\xi_{\|}$ is caused by the parallel diffusion alone. It is reasonable, therefore, to expect that the observed harder rigidity spectra of $\xi_{\bot}$ and $\xi_z$ are due to effects from drift. Based on numerical simulations of particle propagation in turbulent magnetic field, \citet{Min07} has shown that drifts are suppressed by magnetic turbulence, but the suppression sets in at lower turbulence amplitudes for low-energy than for high-energy cosmic rays. This may give a possible explanation for why the contribution of drift streaming results in a harder rigidity spectrum. If this is the case, we may well need two different spectra, representing diffusion and drift, combined in $\xi_{\bot}$ and $\xi_z$,  to reproduce the correct rigidity dependence of the diurnal anisotropy in space.  We will present such analyses elsewhere.\par
Equations (\ref{xparG})-(\ref{xzG}) also imply that the drift contribution to $\xi_{\bot}$ is proportional to $G_{|z|}$, while the drift contribution to $\xi_z$ is proportional to $G_r$. By comparing $G_r$ and $G_{|z|}$ derived from NM and MD data, we find that the rigidity dependences of $G_r$ and $\xi_z$ are harder than those of $G_{|z|}$ and $\xi_{\bot}$. \citet{Yas80} and \citet{Hal94} analyzed the north-south anisotropy observed with NMs and MDs monitoring a wide range of $P_m$ and found $\xi_z$ increasing with the rigidity up to several hundred GV. This is in a qualitative agreement with the present paper.\par
\begin{figure}
\plotone{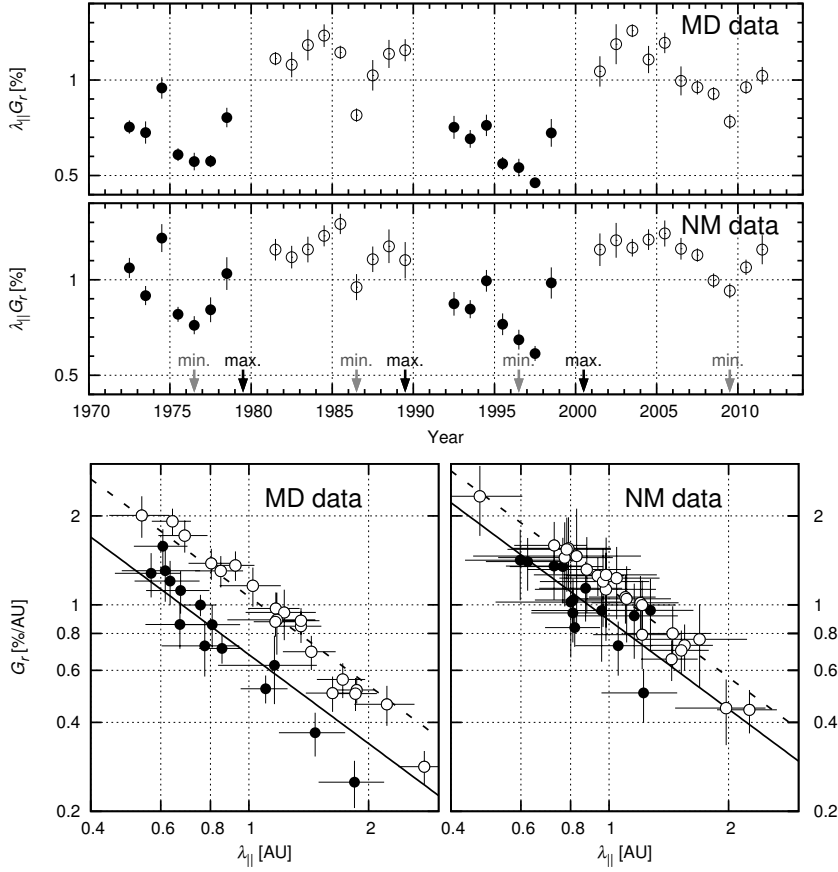}
\caption{Temporal variation of $\lambda_{\|} G_r$ and the correlation between $G_r$ and $\lambda_{\|}$. Upper two panels display yearly mean $\lambda_{\|} G_r$ calculated from $\xi_{\|}/\cos \psi$, each as a function of year. Top panel shows $\lambda_{\|} G_r$ deduced from MD data, while middle panel shows $\lambda_{\|} G_r$ deduced from NM data. Yearly mean values in $A>0$ ($A<0$) epoch are displayed by solid (open) circles, each with an error deduced from the dispersion of monthly values in each year. Data points in years when the polarity reversal is in progress are omitted in this figure. Bottom two panels are scatter plots between $G_r$ and $\lambda_{\|}$ in logarithmic scales derived from MD data (left) and NM data (right). In each panel, yearly mean $G_r$ on the vertical ($y$) axis is plotted as a function of $\lambda_{\|}$ on the horizontal ($x$) axis. Solid and dashed straight lines display the functions $y=c/x$ with a constant parameter $c$ best-fit to data in $A>0$ and $A<0$ epochs, respectively. The solar maximum and minimum periods are indicated by black and gray arrows on the horizontal axis of the middle panel, respectively.}
\label{fig:LG}
\end{figure}

  
We finally discuss the long-term variations of the modulation parameters. Figure \ref{fig:LG} shows the temporal variation of $\lambda_{\|} G_r = \xi_{\|} / \cos \psi$ (see equation (\ref{xparG})). Clearly seen is that the mean magnitude of $\lambda_{\|} G_r$ is significantly smaller in $A>0$ (solid circles) than in $A<0$ periods (open circles). The mean magnitude of $\lambda_{\|} G_r$ derived from MD data and that from NM data in $A<0$ epoch are 1.07$\pm$0.03 and 1.14$\pm$0.02, respectively, which are fairly consistent with each other. The mean magnitudes in $A>0$ periods are 0.68$\pm$0.04 \% and 0.89$\pm$0.05 \%, respectively. Combined with the solar wind convection, this reduction of $\lambda_{\|} G_r$ results in the observed phase shift of the diurnal anisotropy to earlier hours in $A>0$ as suggested by Paper I. We also note that the ratio of $\lambda_{\|} G_r$ for MD to that for NM data is smaller in $A>0$ than in $A<0$ periods, indicating the softer rigidity spectrum of this component for $A>0$ than for $A<0$ (see discussion of figure \ref{fig:IMFcomp} in the preceding section). This larger decrease of $\lambda_{\|} G_r$ in $A>0$ epoch in MD data than in NM data is responsible to the larger phase shift of the diurnal anisotropy in $A>0$ solar minimum epoch in MD data. The harder rigidity spectrum of $\xi_{\bot}$ than that of $\xi_{\|}$ mentioned above is also partly responsible to the larger phase shift in MD data in $A>0$ minimum epochs. \citet{Hal97} used the NM and MD data for analyzing the rigidity spectrum of the diurnal anisotropy and obtained the average $G(p)$ proportional to $p^{-0.1\pm0.2}$ with $P_u = 100\pm25$ GV. Although their spectrum seems to be consistent with $G(p)$ assumed in this paper, such a common spectrum for $\xi_{\|}$ and $\xi_{\bot}$ cannot reproduce the observed feature that the phase shift observed by MD in $A>0$ solar minimum epoch is significantly larger than that by NM.\par
The 11- and 22-year variations are also apparent in the modulation parameters shown in figure \ref{fig:ModParam}. The bidirectional latitudinal density gradient ($G_{|z|}$) in the top panel is positive (negative) in $A>0$ ($A<0$) epoch in accord with the drift model prediction of the local minimum (maximum) of GCR density around the HCS. This 22-year variation looks more significant in MD data than in NM data, with a smaller error of each data point. The mean magnitude of $G_{|z|}$ is larger in $A<0$ than in $A>0$ in both MD and NM data. The 11-year variation is evident in the radial density gradient ($G_r$) in the middle panel of figure \ref{fig:ModParam}, while we cannot identify a clear 22-year variation as reported by \citet{Bie86}. The mean $G_r$ deduced from MD (NM) data is 0.89$\pm$0.11 (1.04$\pm$0.08) \%/AU in $A>0$ epoch, while it is 0.99$\pm$0.12 (1.13$\pm$0.10) \%/AU in $A<0$ epoch. It is noted that we find a poor correlation between temporal variations of $G_{|z|}$ and $G_r$ in both NM and muon data.\par
The mean parallel mean free path ($\lambda_{\|}$), on the other hand, turns out to be significantly larger in the $A<0$ than in the $A>0$ epochs, in the both MD and NM data. We find that the mean $\lambda_{\|}$ deduced from MD (NM) data is 0.90$\pm$0.10 (0.89$\pm$0.06) AU in $A>0$, while it is 1.32$\pm$0.13 (1.14$\pm$0.10) AU in $A<0$. Paper I suggested that the 22-year variation of $\lambda_{\|}$ is responsible for the reduction of $\lambda_{\|} G_r$ in $A>0$ and for the 22-year variation of the diurnal anisotropy. The two bottom panels of figure \ref{fig:LG} show the correlation between $G_r$ and $\lambda_{\|}$ (both in logarithmic scale) on the vertical ($y$) and horizontal ($x$) axes, respectively. Since $\lambda_{\|}$ on the $x$-axis is deduced from $\lambda_{\|} G_r$ divided by $G_r$ on the $y$-axis, data points in this scatter plot align on a straight line when $\lambda_{\|} G_r$ is constant during the analysis period. Solid and dashed straight lines in each panel display functions of $y = c/x$ best-fitting to data in $A>0$ and $A<0$ epochs, respectively, each with the intercept $c$ as a best-fit parameter. It is seen that, for the MD data (left panel) the best-fit $c$ for $A>0$ data (solid circles) is about 64 \% of that for the $A<0$ data (open circles). This is consistent with the lower $\lambda_{\|}$ value derived from MD data for $A>0$ epochs which is 68 \% (=0.90/1.32) of that in $A<0$ epoch, indicating that the 22-year variation of $\lambda_{\|} G_r$ in the left panel is due to the 22-year variation of $\lambda_{\|}$ on the horizontal axis.\par
\begin{figure}
\plotone{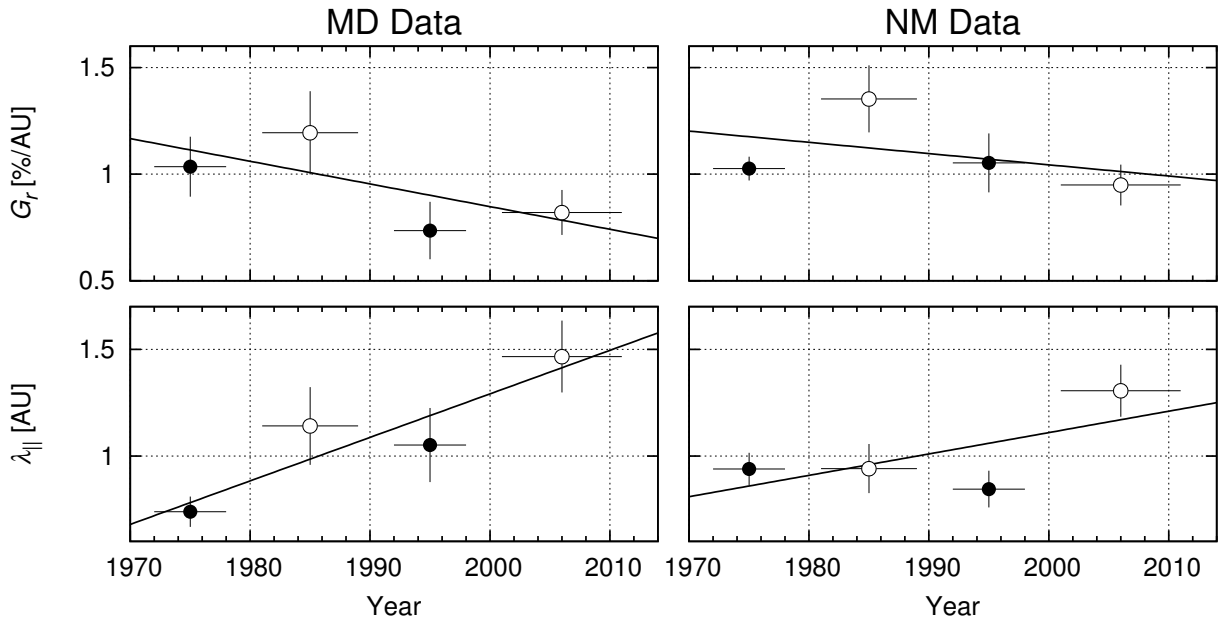}
\caption{Long-term trends of mean $G_r$ and $\lambda_{\|}$ in each solar magnetic polarity epoch. Left (right) two panels display means derived from MD (NM) data. Means in $A>0$ and $A<0$ epochs are plotted by solid and open circles, respectively, at the central year of each epoch. The vertical error is deduced from the dispersion of yearly means in each epoch, while the horizontal bar indicates the period included in each epoch. Solid straight line in each panel displays the linear long-term trend best-fit to four data points.}
\label{fig:CycleAve}
\end{figure}

 
However, as mentioned in connection with figure \ref{fig:ModParam} in the preceding section, we also find that $\lambda_{\|}$s ($G_r$s) from NM and MD data appear to persistently increase (decrease) during the last three solar activity cycles reaching maximum (minimum) in 2008-2009. Figure \ref{fig:CycleAve} displays the mean $G_r$ and $\lambda_{\|}$ in $A>0$ and $A<0$ epochs, each as a functions of time. It is clear particularly in the MD data (left panels) that there is a long-term trends indicated by a best-fit solid line in each panel. This trend enhances the difference between $A>0$ and $A<0$ means of $\lambda_{\|}$, while it reduces the difference between means of $G_r$. The simple means of $G_r$ or $\lambda_{\|}$ in all $A>0$ and $A<0$ epochs are, therefore, seriously biased by these long term trends. If we look at the deviation of each data point from the solid line in the MD data, on the other hand, we find that $G_r$ and $\lambda_{\|}$ are both larger (smaller) in $A<0$ ($A>0$) epoch, although only at one sigma level.\par
The phase-shift of the diurnal anisotropy toward earlier hours in the $A>0$ epochs is a robust consequence of particle drifts in the inhomogeneous large-scale HMF (heliospheric magnetic fields). The observed phase shift in $A>0$ epoch arises naturally in various drift models employing different approaches \citep{Lev76, Erd80, Pot85}. The reproduction of the north-south anisotropy, which is formed by the interplay of drift and perpendicular diffusion, is more challenging for theoretical models. This is particularly true for the $A>0$ epoch, when latitudinal gradients tend to point away from the current sheet, but the intensity minimum of GCRs is not precisely on the HCS. Hence one cannot expect a one-to-one correlation between the field polarity and the NS anisotropy \citep{Oka08}. \citet{Kot01} modeled the 3D anisotropy in a simulation including a wavy HCS with possible variations in the solar wind speed leading to the formation of corotating interaction regions. Their results are in qualitative agreement with the observed phase-shift and reduction of radial gradient in the $A>0$ epochs as well as with the opposite sense of latitudinal gradient around the HCS around solar minima of $A>0$ and $A<0$ epochs. The simulation results for the variation of the NS anisotropy remained inconclusive.\par
It is important to keep in mind that solar cycles are not identical and, as mentioned in the previous section, long-term changes do occur. A particularly interesting recent example is the long and unusual last solar cycle, when the GCR intensity at the Earth reached record-high level \citep{Mew10}. The most plausible explanation is that the magnetic field was weakest ever recorded \citep{McC08} and the weaker field allowed faster diffusion of GCRs into the inner part of the heliosphere. Another remarkable feature of the last solar cycle was that the HCS remained tilted for a long time and did not flatten the same way as in other cycles. Figure \ref{fig:IMFcomp} shows that $\xi _z$ turned out to be larger in the last solar minimum than during previous solar minima. This most likely shows the effect of the tilted HCS. The streaming component normal to the HCS cannot abruptly change, but has to change continuously at the HCS. Hence $\xi _z$ has to go to a small value when the HCS flattens, while it can be larger if the HCS is tilted. This feature is more apparent for MD data than for NM data.\par
The dynamic range of $\lambda_{\|}$ (or $G_r$) due to the 11-year variation in the lower panels of figure \ref{fig:LG} is close to an order of magnitude and much larger than the 22-year variation. Small signature of its 22-year variation can be easily masked by the 11-year variation with much larger amplitude. In order to analyze the 22-year variation of each modulation parameter, therefore, it is necessary to minimize the influence of the 11-year variation as much as possible. Also simple means of $\lambda_{\|}$ and $G_r$ in each of $A>0$ and $A<0$ epochs may be seriously biased by their long term trends as seen above. For identifying the physical origin of the 22-year variation correctly, it is also necessary to analyze its rigidity dependence. The long-term observation with the Nagoya MD, as well as the observations with NMs, makes such analyses possible.

\acknowledgments
This work is supported in part by the joint research program of the Solar-Terrestrial Environment Laboratory (STEL), Nagoya University. The observations with the Nagoya multi-directional muon detector are supported by the Nagoya University. The Bartol Research Institute neutron monitor program, which operates Newark, Thule and McMurdo neutron monitors, is supported by National Science Foundation grant ATM-0000315. We thank the World Data Center for Cosmic Rays, Solar-Terrestrial Environment Laboratory, Nagoya University, for providing the neutron monitor data analyzed in this paper. Wilcox Solar Observatory data used in this study was obtained via the web site \url{http://wso.stanford.edu} at 2014:03:19$\_$01:10:41 PDT courtesy of J.T. Hoeksema. The Wilcox Solar Observatory is currently supported by NASA. JK thanks for the support and hospitality of the STEL and the Shinshu University supplied during his stay as the visiting professor of the STEL.
\clearpage

\appendix
\section{Appendix: Numerical data of anisotropy components and modulation parameters obtained in the present paper}
For readers references, we list in Table \ref{tab:xiMD} numerical data of the anisotropy and modulation parameters derived from MD data at 60 GV. Note that the amplitude and phase in these tables are corrected for the Compton-Getting effect arising from the Earth's orbital motion around the Sun (see section 2.2 in the text). We confirmed that the anisotropy components derived from NMs by us in each year are fairly consistent with the components given in Paper I (see Table 2 in their paper), which analyzed the same NM data in the similar manner during an overlapping period between 1970 and 1988. The amplitude and phase of the common vector derived in our analyses of the MD data are shown in figure \ref{fig:common} (see text). It is seen that the amplitude of the common vector is small and the phase is almost stable around $\sim$06:00 local solar time. 
\clearpage

\section{Appendix: Dependence on the upper limiting rigidity}
\begin{figure}
\plotone{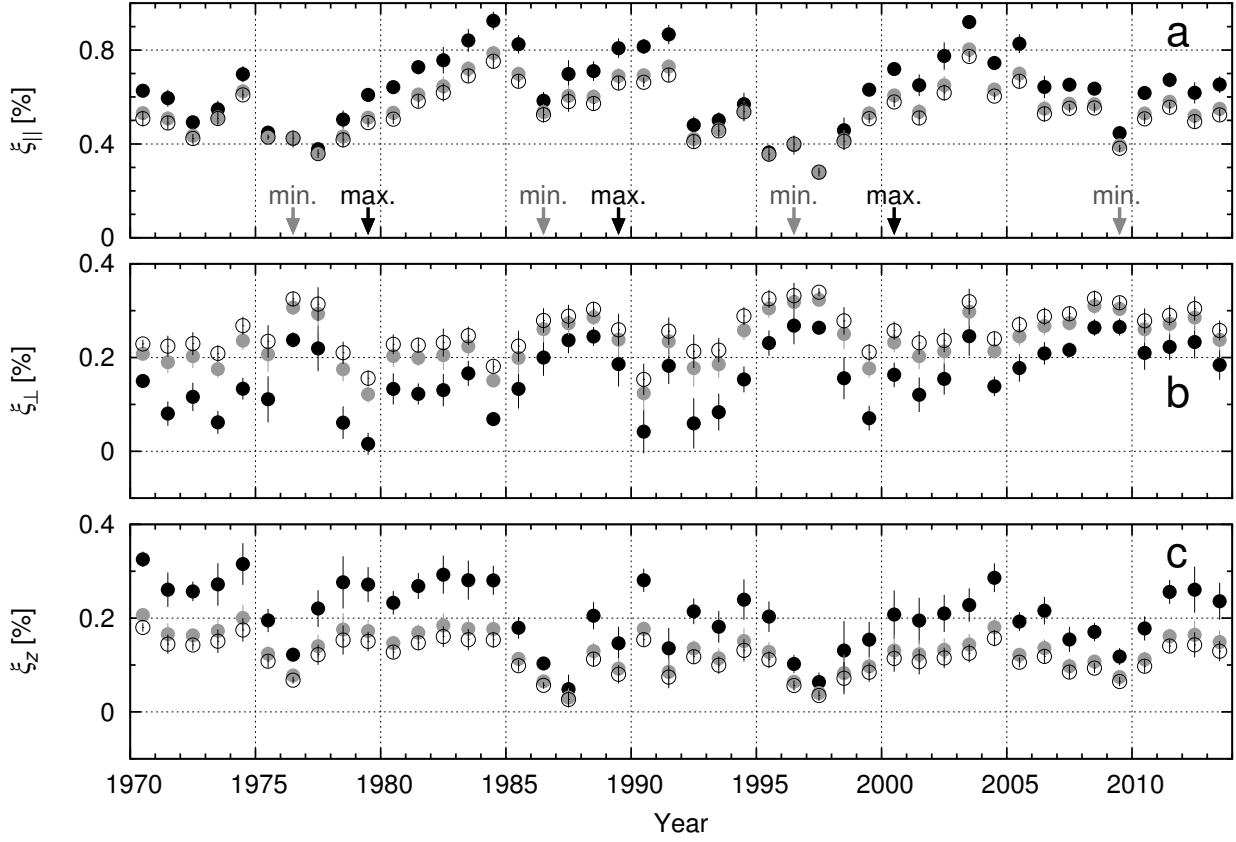}
\caption{Three components of the free-space anisotropy derived from MD data by assuming $P_u=$100, 200, 300 GV. Solid black, solid gray and open circles represent the components obtained with $P_u=$100, 200, 300 GV, respectively. The solar maximum and minimum periods are indicated by black and gray arrows on the horizontal axis of the top panel, respectively.}
\label{fig:PuMD}
\end{figure}

Following analyses in Papers I and II, we assumed in this paper the rigidity spectrum of the anisotropy in equation (9) with $\gamma$ and $P_u$ fixed at 0 and 100 GV respectively. This choice of the spectrum is rather subjective, lacking firm physical or observational proof. In this section, we show how $\beta$ (the ratio between anisotropies and modulation parameters derived from NM and MD data) depends on the upper limiting rigidity ($P_u$) assumed and that our major conclusions on the rigidity dependence derived from $\beta$ are not affected by changing $P_u$. Figure \ref{fig:PuMD} displays anisotropy components derived from MD data with three different $P_u$s. We choose this range of $P_u$ between 100 and 300 GV referring to the solar cycle variation of $P_u$ reported in \citet{Mun97}. We confirmed that the anisotropy derived from NM data is almost insensitive to changing $P_u$ as pointed by Paper II, while the anisotropy derived from MD data changes significantly. The increase of $P_u$ with the same spectral index ($\gamma$) results in the reduction of the amplitude of the free space anisotropy. It also results in the phase of the diurnal anisotropy in free space shifting to earlier hours, due to the reduced average deflection of GCR orbits in the geomagnetic field. Features of anisotropy components in figure \ref{fig:PuMD} changing with $P_u$ are interpreted in terms of these natures of the free space anisotropy. Table \ref{tab:beta} lists mean $\beta$ for three $P_u$s. Firstly, the mean $\beta_{\xi_{\|}}$ close to (or slightly smaller than) one for all $P_u$s indicate $\xi_{\|}$ being similar in NM and MD data, while it is significantly smaller in $A>0$ than in $A<0$ for each $P_u$. Second, the mean $\beta_{\xi_{\bot}}$ and $\beta_{\xi_z}$ are both significantly larger than one indicating harder rigidity spectra of $\xi_{\bot}$ and $\xi_z$ than that of $\xi_{\|}$. The mean $\beta_{\xi_z}$ is always larger than the mean $\beta_{\xi_{\bot}}$. Third, the mean $\beta_{G_{|z|}}$ and $\beta_{G_r}$ are significantly smaller than one for all $P_u$s.

\clearpage

\input{tab1.tex}

\input{tab2.tex}

\input{tab3.tex}

\clearpage
\epsscale{0.8}
\clearpage
\epsscale{0.8}
\clearpage
\epsscale{1}
\begin{figure}
\plotone{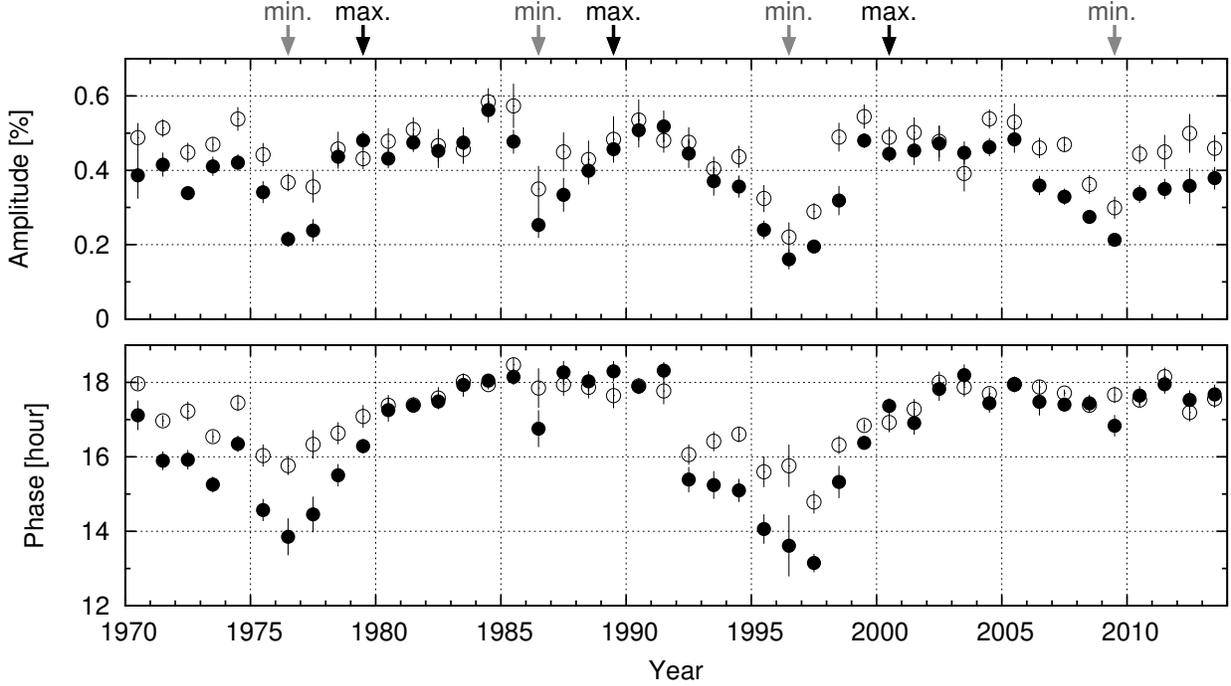}
\caption{Diurnal anisotropy in free-space. Yearly mean amplitude in \% and phase (denoted by the local solar time of the maximum intensity in hour) of the diurnal anisotropy are displayed in the upper and lower panels, respectively, each as a function of year on the horizontal axis. The solid and open circles display the anisotropy obtained from MD data at 60 GV and from NM data at 17 GV, respectively (see Table \ref{tab:xiMD} in Appendix A for numerical data from MD). The diurnal anisotropy in this figure is corrected for the Compton-Getting effect arising from the Earth's orbital motion around the Sun (see text). Yearly mean and error are deduced from the mean and dispersion of monthly values, respectively. The solar maximum and minimum periods are indicated by black and gray arrows above the upper panel, respectively.}
\label{fig:diurnal}
\end{figure}
\clearpage
\clearpage
\clearpage
\clearpage
\clearpage
\epsscale{1}
\clearpage
\epsscale{1}
\begin{figure}
\plotone{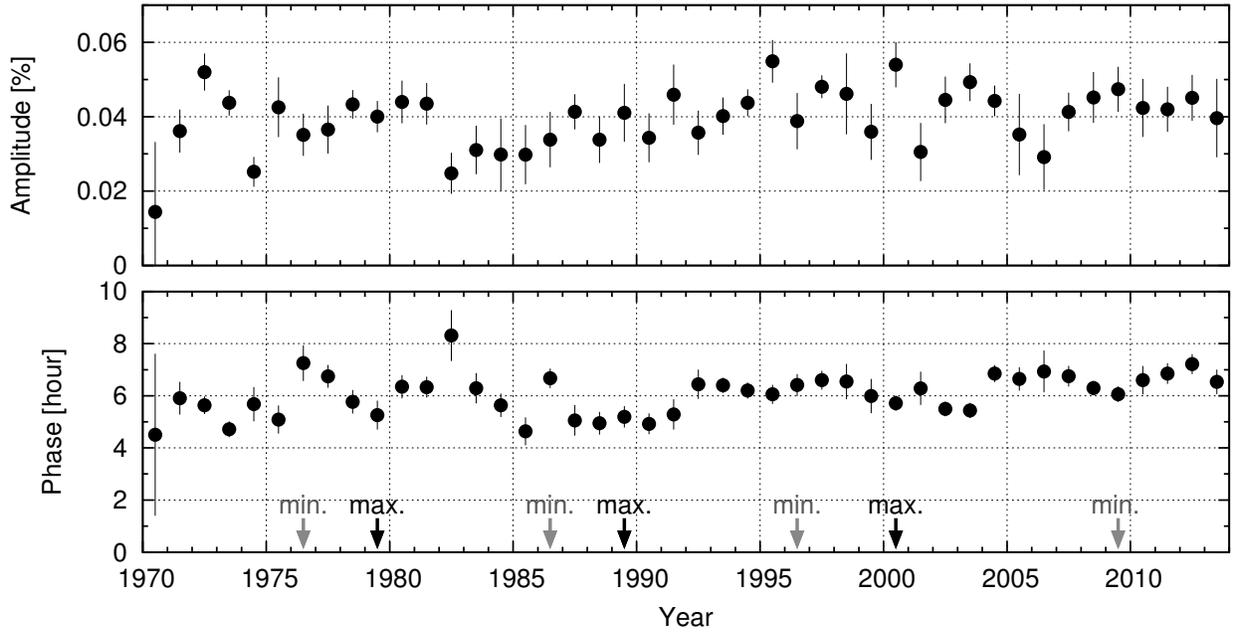}
\caption{Long-term variation of the common vector. Yearly mean amplitude in \% and phase (denoted by the local solar time of the maximum intensity) in hour of the common vector are displayed in the upper and lower panels, respectively, each as a function of year on the horizontal axis. The common vector is introduced as a free parameter representing the atmospheric temperature effect on the diurnal anisotropy observed with the MD (see text). Yearly mean and error are deduced from the mean and dispersion of monthly values, respectively. The solar maximum and minimum periods are indicated by black and gray arrows on the horizontal axis of the lower panel, respectively.}
\label{fig:common}
\end{figure}
\clearpage
\epsscale{1}

\end{document}

%% file: tab1.tex
\begin{deluxetable}{ccrr}
   \tabletypesize{\footnotesize}
   \tablewidth{0pt}
   \tablecaption{Neutron Monitors (NMs) and Muon Detectors (MD) used in this paper}
   \tablehead{\colhead{\shortstack{Detector\\type}}&\colhead{\shortstack{Station\\(geographic lat., long.)}}&\colhead{$P_c$ (GV)}&\colhead{$P_m$ (GV)}}
   \startdata
      & \shortstack{Swarthmore/Newark\\(39.70$^{\circ}$, -75.70$^{\circ}$)} &  2.0 GV & 17.0 GV\\
   NM\tablenotemark{a} & \shortstack{Thule\\(76.60$^{\circ}$, -68.80$^{\circ}$)}             &  0.0 GV & 17.0 GV\\
      & \shortstack{Alert\\(82.50$^{\circ}$, -62.30$^{\circ}$)}             &  0.0 GV & 17.0 GV\\
      & \shortstack{McMurdo\\(-77.95$^{\circ}$, 166.60$^{\circ}$)}          &  0.0 GV & 17.0 GV\\
   \hline
      & \shortstack{Nagoya\\(35.15$^{\circ}$, 139.97$^{\circ}$)}            &         &        \\
   \cline{2-2}
      & Directional channel                             &         &        \\
   \cline{2-4}
      & V                                               & 10.1 GV &  59.4 GV\\
      & N                                               & 10.8 GV &  64.6 GV\\
      & S                                               & 10.0 GV &  62.6 GV\\
      & E                                               & 12.8 GV &  66.7 GV\\
      & W                                               &  9.7 GV &  61.8 GV\\
      & NE                                              & 12.9 GV &  72.0 GV\\
   MD & NW                                              &  9.1 GV &  66.6 GV\\
      & SE                                              & 11.5 GV &  69.3 GV\\
      & SW                                              &  9.5 GV &  65.6 GV\\
      & N2                                              &  8.6 GV &  83.0 GV\\
      & S2                                              &  9.5 GV &  80.5 GV\\
      & E2                                              & 13.2 GV &  88.3 GV\\
      & W2                                              &  8.7 GV &  79.3 GV\\
      & N3                                              &  8.7 GV & 105.0 GV\\
      & S3                                              &  9.5 GV & 103.7 GV\\
      & E3                                              & 17.1 GV & 113.7 GV\\
      & W3                                              &  8.6 GV & 103.0 GV\\
   \enddata
   \tablecomments{The geomagnetic cut-off rigidity ($P_c$) and median primary rigidity ($P_m$) in GV are listed for each directional channel together with the geographic latitude and longitude of the location of the detector in degrees. The Nagoya MD has 17 directional channels, while each NM measures only omni-directional intensity which is regarded as the vertical intensity on average.}
\tablenotetext{a}{For deriving the diurnal anisotropy at 17 GV, we use Swarthmore NM data for a period between 1970 and 1978, while we use Newark NM data for a period between 1979 and 2013. For deriving the north-south anisotropy at 17 GV, we use Thule and McMurdo NMs for two periods, one between 1970 and 1975 and the other between 1979 and 2013, while we use Alert and McMurdo for a period between 1976 and 1978.}
   \label{tab:character}
\end{deluxetable}

%% file: tab2.tex
\begin{deluxetable}{cccccccccc}
   \tabletypesize{\footnotesize}
   \rotate
   \tablewidth{0pt}
   \tablecaption{Anisotropy components and modulation parameters derived from the Nagoya MD data}
   \tablehead{\colhead{year}&\colhead{sgn($A$)\tablenotemark{a}}&\colhead{Amplitude [\%]}&\colhead{Phase [hour]}&\colhead{$\xi_{\|}$ [\%]}&\colhead{$\xi_{\bot}$ [\%]}&\colhead{$\xi_{z}$ [\%]}&\colhead{$G_{|z|}$ [\%/AU]}&\colhead{$G_r$ [\%/AU]}&\colhead{$\lambda_{\|}$ [AU]}}
   \startdata
   1970 & $\ast$ & 0.39$\pm$0.07  & 17.1$\pm$0.4  & 0.63$\pm$0.01 & 0.15$\pm$0.01 &  0.33$\pm$ 0.02 &    $\ast$     & 1.03$\pm$0.05 & 1.02$\pm$0.14\\
   1971 & $\ast$ & 0.42$\pm$0.04  & 15.9$\pm$0.3  & 0.60$\pm$0.04 & 0.08$\pm$0.03 &  0.26$\pm$ 0.04 &    $\ast$     & 1.03$\pm$0.21 & 0.89$\pm$0.19\\
   1972 &   $+$  & 0.34$\pm$0.02  & 15.9$\pm$0.3  & 0.49$\pm$0.03 & 0.12$\pm$0.04 &  0.26$\pm$ 0.03 & 0.32$\pm$0.09 & 1.00$\pm$0.09 & 0.76$\pm$0.08\\
   1973 &   $+$  & 0.41$\pm$0.03  & 15.3$\pm$0.3  & 0.55$\pm$0.04 & 0.06$\pm$0.03 &  0.27$\pm$ 0.05 & 0.17$\pm$0.08 & 1.28$\pm$0.22 & 0.57$\pm$0.11\\
   1974 &   $+$  & 0.42$\pm$0.03  & 16.3$\pm$0.3  & 0.70$\pm$0.04 & 0.13$\pm$0.03 &  0.32$\pm$ 0.05 & 0.43$\pm$0.09 & 1.58$\pm$0.23 & 0.61$\pm$0.10\\
   1975 &   $+$  & 0.34$\pm$0.03  & 14.6$\pm$0.3  & 0.45$\pm$0.02 & 0.11$\pm$0.05 &  0.20$\pm$ 0.03 & 0.27$\pm$0.15 & 0.71$\pm$0.08 & 0.86$\pm$0.11\\
   1976 &   $+$  & 0.21$\pm$0.03  & 13.9$\pm$0.5  & 0.42$\pm$0.04 & 0.24$\pm$0.02 &  0.12$\pm$ 0.02 & 0.61$\pm$0.05 & 0.52$\pm$0.06 & 1.10$\pm$0.15\\
   1977 &   $+$  & 0.24$\pm$0.04  & 14.5$\pm$0.5  & 0.38$\pm$0.03 & 0.22$\pm$0.05 &  0.22$\pm$ 0.04 & 0.59$\pm$0.13 & 0.86$\pm$0.15 & 0.67$\pm$0.13\\
   1978 &   $+$  & 0.44$\pm$0.04  & 15.5$\pm$0.4  & 0.50$\pm$0.04 & 0.06$\pm$0.04 &  0.28$\pm$ 0.06 & 0.20$\pm$0.13 & 1.30$\pm$0.28 & 0.62$\pm$0.14\\
   1979 & $\ast$ & 0.48$\pm$0.03  & 16.3$\pm$0.2  & 0.61$\pm$0.03 & 0.02$\pm$0.03 &  0.27$\pm$ 0.04 &    $\ast$     & 1.38$\pm$0.22 & 0.68$\pm$0.12\\
   1980 & $\ast$ & 0.43$\pm$0.03  & 17.3$\pm$0.4  & 0.64$\pm$0.03 & 0.13$\pm$0.04 &  0.23$\pm$ 0.03 &    $\ast$     & 0.99$\pm$0.11 & 1.06$\pm$0.12\\
   1981 &   $-$  & 0.47$\pm$0.03  & 17.4$\pm$0.2  & 0.73$\pm$0.04 & 0.12$\pm$0.03 &  0.27$\pm$ 0.03 &-0.44$\pm$0.10 & 1.38$\pm$0.17 & 0.80$\pm$0.10\\
   1982 &   $-$  & 0.45$\pm$0.05  & 17.5$\pm$0.3  & 0.76$\pm$0.06 & 0.13$\pm$0.04 &  0.29$\pm$ 0.05 &-0.50$\pm$0.15 & 2.01$\pm$0.33 & 0.54$\pm$0.10\\
   1983 &   $-$  & 0.47$\pm$0.05  & 17.9$\pm$0.4  & 0.84$\pm$0.05 & 0.17$\pm$0.03 &  0.28$\pm$ 0.05 &-0.60$\pm$0.11 & 1.72$\pm$0.22 & 0.69$\pm$0.10\\
   1984 &   $-$  & 0.56$\pm$0.04  & 18.0$\pm$0.2  & 0.92$\pm$0.04 & 0.07$\pm$0.02 &  0.28$\pm$ 0.04 &-0.27$\pm$0.07 & 1.92$\pm$0.20 & 0.64$\pm$0.08\\
   1985 &   $-$  & 0.48$\pm$0.04  & 18.1$\pm$0.3  & 0.82$\pm$0.04 & 0.13$\pm$0.05 &  0.18$\pm$ 0.03 &-0.36$\pm$0.13 & 0.85$\pm$0.11 & 1.35$\pm$0.17\\
   1986 &   $-$  & 0.25$\pm$0.04  & 16.8$\pm$0.5  & 0.58$\pm$0.04 & 0.20$\pm$0.04 &  0.10$\pm$ 0.02 &-0.50$\pm$0.11 & 0.50$\pm$0.07 & 1.62$\pm$0.24\\
   1987 &   $-$  & 0.33$\pm$0.05  & 18.3$\pm$0.4  & 0.70$\pm$0.06 & 0.24$\pm$0.03 &  0.05$\pm$ 0.04 &-0.72$\pm$0.13 & 0.46$\pm$0.08 & 2.22$\pm$0.39\\
   1988 &   $-$  & 0.40$\pm$0.04  & 18.0$\pm$0.3  & 0.71$\pm$0.05 & 0.24$\pm$0.03 &  0.21$\pm$ 0.03 &-0.82$\pm$0.07 & 0.97$\pm$0.13 & 1.17$\pm$0.18\\
   1989 &   $-$  & 0.46$\pm$0.04  & 18.3$\pm$0.3  & 0.81$\pm$0.05 & 0.19$\pm$0.05 &  0.15$\pm$ 0.04 &-0.65$\pm$0.18 & 0.94$\pm$0.19 & 1.23$\pm$0.25\\
   1990 & $\ast$ & 0.51$\pm$0.05  & 17.9$\pm$0.2  & 0.82$\pm$0.04 & 0.04$\pm$0.05 &  0.28$\pm$ 0.03 &    $\ast$     & 1.49$\pm$0.14 & 0.73$\pm$0.08\\
   1991 & $\ast$ & 0.52$\pm$0.05  & 18.3$\pm$0.3  & 0.87$\pm$0.05 & 0.18$\pm$0.04 &  0.14$\pm$ 0.05 &    $\ast$     & 1.02$\pm$0.28 & 1.25$\pm$0.35\\
   1992 &   $+$  & 0.45$\pm$0.04  & 15.4$\pm$0.4  & 0.48$\pm$0.04 & 0.06$\pm$0.06 &  0.21$\pm$ 0.03 & 0.22$\pm$0.22 & 1.12$\pm$0.19 & 0.67$\pm$0.13\\
   1993 &   $+$  & 0.37$\pm$0.04  & 15.2$\pm$0.4  & 0.50$\pm$0.04 & 0.08$\pm$0.04 &  0.18$\pm$ 0.04 & 0.25$\pm$0.14 & 0.86$\pm$0.14 & 0.81$\pm$0.15\\
   1994 &   $+$  & 0.36$\pm$0.03  & 15.1$\pm$0.4  & 0.57$\pm$0.05 & 0.15$\pm$0.03 &  0.24$\pm$ 0.05 & 0.45$\pm$0.09 & 1.20$\pm$0.21 & 0.63$\pm$0.12\\
   1995 &   $+$  & 0.24$\pm$0.03  & 14.1$\pm$0.4  & 0.36$\pm$0.04 & 0.23$\pm$0.03 &  0.20$\pm$ 0.04 & 0.60$\pm$0.08 & 0.73$\pm$0.16 & 0.77$\pm$0.18\\
   1996 &   $+$  & 0.16$\pm$0.03  & 13.6$\pm$0.9  & 0.40$\pm$0.05 & 0.27$\pm$0.04 &  0.10$\pm$ 0.02 & 0.61$\pm$0.08 & 0.37$\pm$0.07 & 1.47$\pm$0.28\\
   1997 &   $+$  & 0.19$\pm$0.02  & 13.1$\pm$0.3  & 0.28$\pm$0.03 & 0.26$\pm$0.02 &  0.06$\pm$ 0.03 & 0.65$\pm$0.04 & 0.25$\pm$0.05 & 1.84$\pm$0.35\\
   1998 &   $+$  & 0.32$\pm$0.04  & 15.3$\pm$0.5  & 0.46$\pm$0.06 & 0.16$\pm$0.05 &  0.13$\pm$ 0.07 & 0.49$\pm$0.15 & 0.62$\pm$0.17 & 1.16$\pm$0.33\\
   1999 & $\ast$ & 0.48$\pm$0.02  & 16.4$\pm$0.2  & 0.63$\pm$0.03 & 0.07$\pm$0.03 &  0.15$\pm$ 0.04 &    $\ast$     & 0.78$\pm$0.15 & 1.27$\pm$0.25\\
   2000 & $\ast$ & 0.44$\pm$0.03  & 17.4$\pm$0.3  & 0.72$\pm$0.03 & 0.16$\pm$0.03 &  0.21$\pm$ 0.06 &    $\ast$     & 1.01$\pm$0.22 & 1.11$\pm$0.24\\
   2001 &   $-$  & 0.45$\pm$0.04  & 16.9$\pm$0.4  & 0.65$\pm$0.05 & 0.12$\pm$0.04 &  0.19$\pm$ 0.05 &-0.33$\pm$0.12 & 0.89$\pm$0.22 & 1.18$\pm$0.30\\
   2002 &   $-$  & 0.47$\pm$0.05  & 17.8$\pm$0.4  & 0.77$\pm$0.06 & 0.15$\pm$0.04 &  0.21$\pm$ 0.04 &-0.57$\pm$0.13 & 1.16$\pm$0.18 & 1.02$\pm$0.18\\
   2003 &   $-$  & 0.45$\pm$0.04  & 18.2$\pm$0.3  & 0.92$\pm$0.03 & 0.25$\pm$0.05 &  0.23$\pm$ 0.04 &-0.88$\pm$0.16 & 1.36$\pm$0.16 & 0.92$\pm$0.11\\
   2004 &   $-$  & 0.46$\pm$0.03  & 17.4$\pm$0.3  & 0.74$\pm$0.04 & 0.14$\pm$0.03 &  0.29$\pm$ 0.04 &-0.41$\pm$0.07 & 1.30$\pm$0.16 & 0.85$\pm$0.12\\
   2005 &   $-$  & 0.48$\pm$0.04  & 17.9$\pm$0.3  & 0.83$\pm$0.05 & 0.18$\pm$0.03 &  0.19$\pm$ 0.02 &-0.51$\pm$0.09 & 0.88$\pm$0.10 & 1.35$\pm$0.16\\
   2006 &   $-$  & 0.36$\pm$0.03  & 17.5$\pm$0.4  & 0.64$\pm$0.05 & 0.21$\pm$0.03 &  0.22$\pm$ 0.03 &-0.45$\pm$0.06 & 0.69$\pm$0.10 & 1.44$\pm$0.22\\
   2007 &   $-$  & 0.33$\pm$0.03  & 17.4$\pm$0.2  & 0.65$\pm$0.03 & 0.22$\pm$0.02 &  0.15$\pm$ 0.03 &-0.43$\pm$0.04 & 0.51$\pm$0.06 & 1.87$\pm$0.20\\
   2008 &   $-$  & 0.27$\pm$0.02  & 17.4$\pm$0.3  & 0.64$\pm$0.03 & 0.26$\pm$0.02 &  0.17$\pm$ 0.02 &-0.49$\pm$0.04 & 0.50$\pm$0.07 & 1.85$\pm$0.25\\
   2009 &   $-$  & 0.21$\pm$0.02  & 16.8$\pm$0.3  & 0.45$\pm$0.02 & 0.26$\pm$0.02 &  0.12$\pm$ 0.02 &-0.44$\pm$0.04 & 0.28$\pm$0.04 & 2.76$\pm$0.39\\
   2010 &   $-$  & 0.34$\pm$0.03  & 17.6$\pm$0.3  & 0.62$\pm$0.03 & 0.21$\pm$0.04 &  0.18$\pm$ 0.03 &-0.43$\pm$0.07 & 0.56$\pm$0.08 & 1.72$\pm$0.24\\
   2011 &   $-$  & 0.35$\pm$0.03  & 17.9$\pm$0.3  & 0.67$\pm$0.04 & 0.22$\pm$0.03 &  0.26$\pm$ 0.03 &-0.53$\pm$0.08 & 0.88$\pm$0.10 & 1.17$\pm$0.14\\
   2012 & $\ast$ & 0.36$\pm$0.05  & 17.5$\pm$0.3  & 0.62$\pm$0.05 & 0.23$\pm$0.04 &  0.26$\pm$ 0.05 &    $\ast$     & 0.93$\pm$0.17 & 1.12$\pm$0.21\\
   2013 & $\ast$ & 0.38$\pm$0.04  & 17.7$\pm$0.3  & 0.65$\pm$0.04 & 0.18$\pm$0.04 &  0.24$\pm$ 0.04 &    $\ast$     & 0.76$\pm$0.14 & 1.34$\pm$0.25\\
   \enddata
\tablecomments{The amplitude and phase (the local solar time of maximum intensity) of the space harmonic vector in figure \ref{fig:diurnal}, three components ($\xi_{\|}$, $\xi_{\bot}$, $\xi_{z}$) of the anisotropy in the solar wind frame in figure \ref{fig:IMFcomp} and modulation parameters ($G_{|z|}$, $G_r$, $\lambda_{\|}$) in figure \ref{fig:ModParam}, all derived from the Nagoya MD data, are listed for each year. Yearly mean value and error are deduced from the average and dispersion of monthly values, respectively.} \tablenotetext{a}{Each character in the column ``sgn($A$)'' indicates the polarity of the large-scale solar magnetic field assigned by us referring to the Solar Polar Field Strength available at the WSO web-site; ``$+$'' for a year in $A>0$ epoch, ``$-$'' for a year in $A<0$ epoch and ``$\ast$'' for a year when the polarity reversal is in progress (see text).}
   \label{tab:xiMD}
\end{deluxetable}

%% file: tab3.tex
\begin{deluxetable}{ccccc}
   \tabletypesize{\footnotesize}
   \tablewidth{0pt}
   \tablecaption{Mean $\beta$ values obtained with three different $P_u$s in equation (\ref{G})}
   \tablehead{\colhead{}&\colhead{polarity}&\colhead{$P_u=100$ GV}&\colhead{$P_u=200$ GV}&\colhead{$P_u=300$ GV}}
   \startdata
			  & $A>0$   & 0.77$\pm$0.07 & 0.74$\pm$0.06 & 0.75$\pm$0.06\\
   $\beta_{\xi_{\|}}$     & $A<0$   & 0.94$\pm$0.05 & 0.84$\pm$0.04 & 0.82$\pm$0.04\\
			  & mean & 0.89$\pm$0.05 & 0.81$\pm$0.04 & 0.79$\pm$0.04\\
   \hline
			  & $A>0$   & 1.65$\pm$0.35 & 2.24$\pm$0.34 & 2.36$\pm$0.33\\
   $\beta_{\xi_{\bot}}$   & $A<0$   & 1.26$\pm$0.14 & 1.59$\pm$0.14 & 1.70$\pm$0.14\\
			  & mean & 1.35$\pm$0.14 & 1.80$\pm$0.13 & 1.93$\pm$0.13\\
   \hline
			  & $A>0$   & 4.45$\pm$0.61 & 2.81$\pm$0.39 & 2.45$\pm$0.34\\
   $\beta_{\xi_z}$        & $A<0$   & 6.08$\pm$0.96 & 3.82$\pm$0.61 & 3.32$\pm$0.53\\
			  & mean & 5.22$\pm$0.55 & 3.29$\pm$0.35 & 2.86$\pm$0.31\\
   \hline
			  & $A>0$   & 0.48$\pm$0.10 & 0.68$\pm$0.09 & 0.72$\pm$0.09\\
   $\beta_{G_{|z|}}$      & $A<0$   & 0.35$\pm$0.05 & 0.46$\pm$0.05 & 0.49$\pm$0.06\\
			  & mean & 0.39$\pm$0.05 & 0.53$\pm$0.05 & 0.57$\pm$0.05\\
   \hline
			  & $A>0$   & 0.85$\pm$0.12 & 0.56$\pm$0.08 & 0.50$\pm$0.07\\
   $\beta_{G_r}$          & $A<0$   & 0.87$\pm$0.13 & 0.58$\pm$0.09 & 0.52$\pm$0.08\\
			  & mean & 0.86$\pm$0.08 & 0.57$\pm$0.05 & 0.51$\pm$0.05\\
   \hline
			  & $A>0$   & 1.00$\pm$0.13 & 1.44$\pm$0.18 & 1.62$\pm$0.20\\
   $\beta_{\lambda_{\|}}$ & $A<0$   & 1.16$\pm$0.15 & 1.53$\pm$0.19 & 1.65$\pm$0.21\\
			  & mean & 1.08$\pm$0.09 & 1.47$\pm$0.12 & 1.61$\pm$0.13\\
   \enddata
   \tablecomments{The $\beta$ value is the ratio of the parameter derived from MD data at 60 GV to that derived from NM data at 17 GV (see text). Mean $\beta$ values in $A>0$ and $A<0$ epochs and in the total period consisting of all $A>0$ and $A<0$ epochs are listed. Mean value and error are deduced from the average and dispersion of yearly values.}
   \label{tab:beta}
\end{deluxetable}